\documentstyle[11pt,aaspp4,psfig,flushrt]{article} 
\received{RECEIPT DATE 1996}
\accepted{ACCEPT DATE 1996}
\journalid{VOL}{JOURNAL DATE 1996}
\articleid{START PAGE}{END PAGE}

       \font\sevenrm=cmr7          
\def\split{\gamma\to\gamma\gamma}
\def\erg{\varepsilon}
\def\lambar{\lambda\llap {--}}
\def\teq#1{$\, #1\,$}                         
\def\dover#1#2{\hbox{${{\displaystyle#1 \vphantom{(} }\over{
   \displaystyle #2 \vphantom{(} }}$}}
\begin{document}
\newcommand{\figureout}[3]{\psfig{figure=#1,width=5.5in,angle=#2} 
  \figcaption{#3} } 
\title{PHOTON SPLITTING CASCADES IN GAMMA-RAY PULSARS\\
           AND THE SPECTRUM OF PSR1509-58}

\author{Alice K. Harding, Matthew G. Baring\altaffilmark{1}}
\affil{Laboratory for High Energy Astrophysics \\
       NASA/Goddard Space Flight Center \\
       Greenbelt, MD 20771 \\
       harding@twinkie.gsfc.nasa.gov, baring@lheavx.gsfc.nasa.gov}
\and
\author{Peter L. Gonthier}
\affil{Department of Physics \\
       Hope College \\
       Holland, MI 49423 \\
       gonthier@physics.hope.edu}


\altaffiltext{1}{Compton Fellow, Universities Space Research Association}

\begin{abstract}

Magnetic photon splitting \teq{\gamma\to\gamma\gamma}, a QED process
that becomes important only in magnetic fields approaching the quantum
critical value, $B_{\rm cr} = 4.41 \times 10^{13}$ Gauss, is
investigated as a mechanism for attenuation of  $\gamma$-rays emitted
near the surface of strongly-magnetized pulsars.  Since splitting has no
threshold, it can attenuate photons and degrade their energies below
the threshold for  one-photon pair production, and in high enough fields
it may dominate photon attenuation above pair threshold.  We model
photon splitting attenuation and subsequent splitting cascades in
$\gamma$-ray pulsars, including the  dipole field and curved spacetime
geometry of the neutron star magnetosphere.  We focus  specifically on
PSR1509-58, which has the highest surface magnetic field of all the
$\gamma$-ray pulsars ($B_0 = 3 \times 10^{13}$ Gauss).   We find that
splitting will not be important for most $\gamma$-ray pulsars, i.e.
those with  $B_0 \lesssim 0.2B_{\rm cr}$, either in competition with
pair  production attenuation in pair cascades, or in photon escape
cutoffs in  the spectrum. Photon splitting \it will \rm be important for
$\gamma$-ray pulsars having $B_0 \gtrsim 0.3B_{\rm cr}$, where the
splitting attenuation lengths and  escape energies become comparable to
or less than those for pair production. We compute Monte Carlo spectral
models for PSR1509-58, assuming that either a full photon splitting
cascade or a combination of splitting and pair production (depending on
which splitting modes operate) attenuate a power-law input spectrum.  We
find that photon splitting, or combined splitting and pair production,
can explain the  unusually low cutoff energy (between 2 and 30 MeV) of
PSR1509-58, and that the model cascade spectra, which display strong
polarization, are consistent with the observed spectral points and upper
limits for polar cap emission at a range of magnetic colatitudes up to
$\sim 25^\circ$. 

\end{abstract}


\received{}
\revised{}
\accepted{}

\section{INTRODUCTION}

The discovery of at least five new $\gamma$-ray pulsars by the Compton
Gamma-Ray Observatory (CGRO) and ROSAT has re-ignited theoretical work on
the physical processes and modeling of high-energy radiation from pulsars.
Including the previously known $\gamma$-ray pulsars, Crab (Nolan et al. 1993)
and Vela (Kanbach et al. 1994), the recent detection of pulsed $\gamma$-rays
from PSR B1706-44 (Thompson et al. 1992), Geminga (Halpern \& Holt 1992,
Bertsch et al. 1992), PSR1509-58 (Wilson et al. 1993), PSR B1055-52
(Fierro et al. 1993), PSR B1951+32 (Ramanamurthy et al. 1995) and possibly
PSR0656+14 (Ramanamurthy et al. 1996) bring the total to at least seven.
For the first time, it is possible to look for the similarities and patterns in
the $\gamma$-ray emission characteristics that may reveal clues to the origin
of this emission.  Examples are a preponderance of double pulses, a 
$\gamma$-ray luminosity vs. polar cap current correlation, a spectral
hardness vs. characteristic age correlation, and spectral cutoffs above a
few GeV (e.g. Thompson 1996).  PSR1509-58 stands out among the known
$\gamma$-ray pulsars as having both an unusually low spectral cutoff energy
(somewhere between 2 and 30 MeV) and the highest inferred surface magnetic
field ($3 \times 10^{13}$ Gauss).  It has been detected by the CGRO instruments
operating only in the lowest energy bands, BATSE (Wilson et al. 1993)
and OSSE (Matz et al. 1994), with the higher energy instruments, COMPTEL
(Bennett et al. 1993) and EGRET (Nel et al. 1996), giving upper limits that 
require a cutoff or turnover between 2 and 30 MeV.  There is no evidence
for pulsed TeV emission (Nel et al. 1992).

There are currently two types of models for $\gamma$-ray pulsars being
investigated in detail.  Polar cap models assume that particles are 
accelerated along open field lines near the neutron star by strong parallel 
electric fields (e.g. Arons 1983).  The primary particles induce 
electromagnetic cascades through the creation of electron-positron pairs
by either curvature radiation (Daugherty \& Harding 1982, 1994, 1996) or
inverse-Compton radiation (Sturner \& Dermer 1994) $\gamma$-rays.  
Outer gap models assume that the primary particles are accelerated along
open field lines in the outer magnetosphere, near the null charge surface,
where the corotation charge changes sign, and where strong electric fields
may develop (Cheng, Ho \& Ruderman 1986a,b; Chiang \& Romani 1992;
Romani \& Yadigaroglu 1995).  Since the magnetic fields in the outer gaps
are too low to sustain one-photon pair production cascades, these models
must rely on photon-photon pair production of $\gamma$-rays, interacting
with either non-thermal X-rays from the gap or thermal X-rays from the
neutron star surface, to initiate pair cascades.

Magnetic one-photon pair production, $\gamma\to e^+e^-$, has so far been
the only photon  attenuation mechanism assumed to operate in polar cap
cascade models. Another attenuation mechanism, photon splitting
\teq{\split}, will  also operate in the high-field regions near pulsar
polar caps but has not yet been included in polar cap model
calculations. The rate of photon splitting increases rapidly with
increasing field strength (Adler 1971), so that it may even be the
dominant attenuation process in the highest field pulsars.  There are
several potentially important  consequences of photon splitting for
$\gamma$-ray pulsar models.  Since photon splitting has no threshold, it
can attenuate photons below the threshold for pair production, $\erg =
2/\sin\theta_{\rm kB}$, and can thus produce cutoffs in the spectrum at
lower energies.  Here $\theta_{\rm kB}$ is the angle between the photon
momentum and the magnetic field vectors, and $\erg$ is (hereafter)
expressed in units of $m_ec^2$.  When the splitting rate becomes large
enough, splitting can take place during a photon's propagation through
the neutron star magnetosphere \it before \rm the pair production
threshold is crossed (i.e. before an angle \teq{\sim 2/\erg} to the
field is achieved).  Consequently, the production of secondary electrons
and positrons in pair cascades will be suppressed.  Instead of {\it
pair} cascades, one could have  {\it splitting} cascades, where the
high-energy photons split repeatedly until they escape the
magnetosphere.   The potential importance of photon splitting in neutron
star applications was suggested by Adler (1971), Mitrofanov \it et al.
\rm (1986) and Baring (1988).  Its attenuation and reprocessing
properties have been explored in the contexts of annihilation line
suppression in gamma-ray pulsars (Baring 1993), and spectral formation
of gamma-ray bursts from neutron stars (Baring 1991).   Photon splitting
cascades have also been investigated in models of soft $\gamma$-ray
repeaters, where they will soften the photon spectrum very efficiently
with no production of pairs (Baring 1995, Baring \& Harding 1995a,
Harding \& Baring 1996, Chang et al. 1996).  

In this paper we examine the importance of photon splitting in
$\gamma$-ray pulsar polar cap models (it presumably will not operate in
the low fields of outer gap models).  Following a brief discussion of
the physics of photon splitting in Section~2, we present calculations of
the splitting attenuation lengths and escape energies in the dipole
magnetic field of a neutron star. A preliminary study (Harding, Baring
\& Gonthier 1996) has shown that  splitting will be the primary mode of
attenuation of $\gamma$-rays emitted  parallel to a magnetic field
\teq{B \gtrsim 0.3 B_{\rm cr} = 1.3\times 10^{13}} Gauss.  We then
present, in Section~3, photon splitting cascade models for two cases:
(1) when only one mode of splitting ($\perp \rightarrow
\parallel\parallel$) allowed by the kinematic selection rules (Adler
1971, Shabad 1975) operates, suppressing splitting of photons of
parallel polarization (so that they can only pair produce), but still
permitting photons of perpendicular polarization to either split once or
produce pairs, and (2) when the three splitting modes allowed by CP
(charge-parity) invariance operate, producing mode switching and a
predominantly photon splitting cascade.  In Section~4, model cascade
spectra are compared to the  observed spectrum of PSR1509-58 to
determine the range of magnetic colatitude emission points (if any) that
can produce a spectral cutoff consistent with the data.  These spectra
have cutoff energies that are decreasing functions of the magnetic
colatitude. It is found that a reasonably broad range of polar cap sizes
will accommodate the data, and that strong polarization signatures
appear in the spectra due to the action of photon splitting.

\section{PHOTON SPLITTING AND PAIR CREATION ATTENUATION}

The basic features of magnetic photon splitting \teq{\gamma\to\gamma\gamma} and
the more familiar process of single-photon pair creation \teq{\gamma\to e^+e^-}
are outlined in the next two subsections before investigating their role as
photon attenuation mechanisms in pulsar magnetospheres. Note that throughout
this paper, energies will be rendered dimensionless, for simplicity, using the
scaling factor \teq{m_ec^2}.  Magnetic fields will also often be scaled by the
critical field \teq{B_{\rm cr}}; this quantity will be denoted by a prime:
\teq{B'=B/B_{\rm cr}}.

\subsection{Photon Splitting Rates}

The splitting of photons in two in the presence of a strong magnetic
field is an exotic and comparatively recent prediction of quantum
electrodynamics (QED), with the first correct calculations of the
reaction rate being performed in the early '70s (Bialynicka-Birula and
Bialynicki-Birula 1970; Adler \it et al. \rm 1970; Adler 1971).  Its
relative obscurity to date (compared, for example, with magnetic pair
creation) in the astrophysical community stems partly from the
mathematical complexity inherent in the computation of the rate.  
Splitting is a third-order QED process with a triangular Feynman
diagram.  Hence, though splitting is kinematically possible, when
\teq{B=0}  it is forbidden by a charge conjugation symmetry of QED known 
as Furry's theorem (e.g. see Jauch and Rohrlich, 1980), which states
that ring diagrams that have an odd number of vertices with only
external photon lines generate interaction matrix elements that are
identically zero.  This symmetry is broken by the presence of an
external field.  The splitting of photons is  therefore a purely quantum
effect, and has appreciable reaction rates only when the magnetic field
is at least a significant fraction of the quantum critical field
\teq{B_{\rm cr} = m_e^2c^3/(e\hbar )=4.413\times 10^{13}} Gauss.
Splitting into more than two photons is prohibited in the limit of zero
dispersion because of the lack of available quantum phase space 
(Minguzzi 1961).

The reaction rate for splitting is immensely complicated by dispersive
effects (e.g. Adler 1971; Stoneham 1979) caused by the deviation of the
refractive index from unity in the strong field.  Consequently, 
manageable expressions for the rate of splitting are only possible in
the limit of zero dispersion, and are still then complicated triple
integrations (see Stoneham 1979, and also Ba\u{\i}er, Mil'shte\u{\i}n,
and Sha\u{\i}sultanov 1986 for electric field splitting) due to the
presence of magnetic electron propagators in the matrix element.  Hence
simple expressions for the rate of splitting of a photon of energy
\teq{\omega} in a field \teq{B} were first obtained by Bialynicka-Birula
and  Bialynicki-Birula (1970), Adler \it et al. \rm (1970) and Adler
(1971) in the low-energy, non-dispersive limit: \teq{\omega B/B_{\rm
cr}\lesssim 1}. The total rate in this limit, averaged over photon
polarizations (Papanyan and Ritus, 1972), is expressible in terms of an
attenuation coefficient
\begin{equation}
   T_{\rm sp} (\omega )\;\approx\; {{\alpha^3}\over{10\pi^2}}\,         
   \dover{1}{\lambar}\, {\biggl({{19}\over{315}}\biggr) }^2\, 
   B'^6\, {\cal C}(B')\, \omega^5\,\sin^6\theta_{\rm kB} \quad ,
  \label{eq:splitotrate}
\end{equation}
where \teq{\alpha =e^2/\hbar c\approx 1/137} is the fine structure
constant, \teq{\lambar =\hbar /(m_ec)} is the Compton wavelength of the
electron, and \teq{\theta_{\rm kB}} is the angle between the photon
momentum and the magnetic field vectors.  Here \teq{{\cal C}(B')} is a
strong-field modification factor (derivable, for example, from Eq.~41 of
Stoneham, 1979: see Eq.~[\ref{eq:splitratecorr}] below) that
approximates unity when \teq{B\ll B_{\rm cr}} and scales as \teq{B^{-6}}
for \teq{B\gg B_{\rm cr}}.

The corresponding differential spectral rate for the splitting of photons of
energy \teq{\omega} (with \teq{\omega\ll 1}) into photons of energies 
\teq{\omega'} and \teq{\omega -\omega'} is
\begin{equation}
   T_{\rm sp}(\omega ,\omega')\;\approx\; 30\,\dover{\omega'^2
   (\omega -\omega' )^2}{\omega^5}\, T_{\rm sp}(\omega )\quad .
  \label{eq:splitdiffrate}
\end{equation}
Equations~(\ref{eq:splitotrate}) and~(\ref{eq:splitdiffrate}) are valid
(Baring, 1991) when \teq{\omega B' \sin\theta_{\rm kB}\lesssim 1}, which
for pulsar fields and \teq{\omega \sin\theta_{\rm kB} \lesssim 2}, generally corresponds to the regime of weak vacuum dispersion.
Reducing \teq{\theta_{\rm kB}} or
\teq{B} dramatically increases  the photon energy required for splitting
to operate in a neutron star environment.  The produced photons emerge
at an angle \teq{\theta_{\rm kB}} to the field since splitting is a
collinear process in the low-dispersion limit.

Adler (1971) observed that in the low-energy limit, the splitting rate
was strongly dependent on the polarization states of the initial and
final photons; this feature prompted the suggestion by Adler et al.
(1970) and Usov and Shabad (1983) that photon splitting  should be a
powerful polarizing mechanism in pulsars.  The polarization-dependent
rates can be taken from Eq.~(23) of Adler (1971), which can be related
to equations~(\ref{eq:splitotrate}) or~(\ref{eq:splitdiffrate}) via
\begin{equation}
   T^{\rm sp}_{\perp\to\parallel\parallel}\; =\; 
   \dover{1}{2}\, T^{\rm sp}_{\parallel\to\perp\parallel}\; =\; 
   \biggl( \dover{{\cal M}_1^2}{{\cal M}_2^2} {\biggr)}^2\,
   T^{\rm sp}_{\perp\to\perp\perp}\; =\; \dover{2 {\cal M}_1^2\,
   T_{\rm sp} }{ 3{\cal M}_1^2 + {\cal M}_2^2 } \quad ,
  \label{eq:splitpolrate}
\end{equation}
where the scattering amplitude coefficients 
\begin{eqnarray}
   {\cal M}_1 & = &\dover{1}{B'^4}\int^{\infty}_{0} \dover{ds}{s}\, e^{-s/B'}\,
   \Biggl\{ \biggl(-\dover{3}{4s}+\dover{s}{6}\biggr)\,\dover{\cosh s}{\sinh s} 
   +\dover{3+2s^2}{12\sinh^2s}+\dover{s\cosh s}{2\sinh^3s}\Biggr\}\nonumber\\
   {\cal M}_2 & = &\dover{1}{B'^4}\int^{\infty}_{0} \dover{ds}{s}\, e^{-s/B'}\,
   \Biggl\{ \dover{3}{4s}\,\dover{\cosh s}{\sinh s} +
   \dover{3-4s^2}{4\sinh^2s} - \dover{3s^2}{2\sinh^4s}\Biggr\}
  \label{eq:splitcoeff}
\end{eqnarray}
are given in Adler (1971) and Eq.~41 of Stoneham (1979).  In the limit
of \teq{B\ll B_{\rm cr}}, \teq{{\cal M}_1\approx 26/315} and \teq{{\cal
M}_2\approx 48/315}, while in the limit of \teq{B\gg B_{\rm cr}},
equation~(\ref{eq:splitcoeff}) produces \teq{{\cal M}_1\approx
1/(6B'^3)} and \teq{{\cal M}_2\approx 1/(3B'^4)}. The factor of two in
the numerator of the right hand side of 
equation~(\ref{eq:splitpolrate}) accounts for the duplicity of photons
produced in splitting. The photon polarization labelling convention of
Stoneham (1979) is adopted here (this standard form was not used by
Adler, 1971): the label \teq{\parallel} refers to the state with the
photon's \it electric \rm field vector parallel to the plane containing
the magnetic field and the photon's momentum vector, while \teq{\perp}
denotes the photon's electric field vector being normal to this plane.
Summing over the polarization modes yields the relationship for  the
strong-field modification factor in equation~(\ref{eq:splitotrate}):
\begin{equation}
   {\cal C}(B')\; =\;\dover{1}{12}\,
   {\biggl({{315}\over{19}}\biggr) }^2\, \Bigl( 3{\cal M}_1^2 + {\cal M}_2^2
   \Bigr) \quad .
  \label{eq:splitratecorr}
\end{equation}
Note that, in the absence of vacuum dispersion, the splitting modes
\teq{\perp\to\perp\parallel}, \teq{\parallel\to\perp\perp} and 
\teq{\parallel\to\parallel\parallel} are forbidden by arguments of CP 
(charge-parity) invariance in QED (Adler 1971); dispersive effects admit 
the possibility of non-collinear photon splitting so that there is a 
small but non-zero probability for the \teq{\perp\to\perp\parallel} 
channel.  Equations~(\ref{eq:splitotrate})--(\ref{eq:splitratecorr})
define the rates to be used in the analyses of this paper, and are valid
for \teq{\omega B'\sin\theta_{\rm kB}\ll 1}.  The triple
integral expressions that Stoneham (1979) derives are valid (below pair
creation threshold) for a complete range (i.e. 0 to \teq{\infty}) of the
expansion parameter \teq{\omega B'\sin\theta_{\rm kB}}, but
are not presently in a computational form suitable for use here.  Work
is in progress to address this deficiency (Baring \& Harding 1996), and preliminary results indicate that
equations~(\ref{eq:splitotrate})--(\ref{eq:splitratecorr}) approximate
Stoneham's (1979) formulae to better than two percent for \teq{\omega
B'\sin\theta_{\rm kB}\leq 0.2}, and differs by at most a
factor of around 2.5 for \teq{\omega B'\sin\theta_{\rm kB}
\sim 1.5}, the value relevant to the calculations of this paper; 
the splitting rate given by Stoneham's formulae initially increase 
above the low energy limits as \teq{\omega B'\sin\theta_{\rm kB}} increases.

Recently there has appeared a new result on the rates of photon 
splitting.  Mentzel, Berg \& Wunner (1994) presented an S-matrix
calculation of the rates for the three polarization modes permitted by
CP  invariance that are considered here.  While their formal development
is comparable to an earlier S-matrix formulation of splitting in 
Melrose \& Parle (1983a,b), their presentation of numerical results
appeared to be in violent disagreement (see also their astrophysical 
presentation in Wunner, Sang \& Berg 1995) with the splitting results
obtained via the Schwinger proper-time technique by Adler (1971) and
Stoneham (1979) that comprise
equations~(\ref{eq:splitotrate})--(\ref{eq:splitratecorr}) here. These
results have now been retracted, the disagreement being due to a sign error
in their numerical code (Wilke \& Wunner 1996).  The revised results are in 
much better agreement with the rates computed by Adler (1971). 
However, the revised numerical splitting rates of Wilke \& Wunner (1996) still 
differ by as much as a factor of 3 from Baring \& Harding's (1996)
computations of Stoneham's (1979) general formulae.  
Ba\u{\i}er, Mil'shte\u{\i}n,
\& Sha\u{\i}sultanov (1996) generate numerical results from their
earlier alternative proper-time calculation (Ba\u{\i}er, Mil'shte\u{\i}n,
\& Sha\u{\i}sultanov 1986) that are in accord with Stoneham's and 
Adler's (1971) results and also with those of Baring \& Harding (1996). The
numerical computation of the S-matrix formalism is a formidable task. 
The proper-time analysis, though difficult, is more amenable, and has
been reproduced in the limit of \teq{B\ll B_{\rm cr}} by numerous
authors.  As the S-matrix and
proper-time techniques should produce equivalent results, and indeed
have done so demonstrably in the case of pair production (see DH83 and
Tsai \& Erber 1974), we choose to use the amenable proper-time results
outlined above in the calculations of this paper. 

The above results ignore the fact that the magnetized vacuum is
dispersive and birefringent, so that the phase velocity of light is less
than \teq{c} and depends on the photon polarization.  Dispersion can
therefore alter the kinematics of QED processes such as splitting (Adler
1971), and further dramatically complicates the formalism for the rates
(Stoneham 1979).  Extensive discussions of dispersion in a magnetized
vacuum are presented by Adler (1971) and Shabad (1975); considerations
of plasma dispersion are not relevant to the problem of gamma-ray
emission from pulsars because they become significant only for densities
in excess of around \teq{10^{27}}cm$^{-3}$, which are only attained at
the stellar surface.   Adler (1971) showed that in the limit of \it weak
\rm vacuum dispersion (roughly delineated by $B'\sin\theta_{\rm kB}
\lesssim 1$),  where the refractive indices for the polarization states
are {\it very} close to unity, energy and momentum could be simultaneously
conserved only for the splitting mode \teq{\perp\to\parallel\parallel}
(of the modes permitted by CP invariance) below pair production
threshold.  This kinematic selection rule was demonstrated for subcritical 
fields, where the dispersion is very weak, a regime that generally applies
to gamma-ray pulsar scenarios.  Therefore, it is probable that only the
one mode (\teq{\perp\to\parallel\parallel}) of splitting operates in
gamma-ray pulsars.  This result may be modified by subtle effects such
as those incurred by field non-uniformity.  We adopt a dual scenario in
this paper for the sake of completeness: one in which all CP-permitted
modes of splitting operate, and one in which Adler's kinematic selection
rules are imposed.  Note that in magnetar models of soft gamma repeaters
(e.g Baring 1995, Harding and Baring 1996), where supercritical fields
are employed, moderate vacuum dispersion arises.  In such a regime, it is
not clear whether Adler's selection rules still endure, since his
analysis implicitly uses weak dispersion limits of linear vacuum
polarization results (e.g. see Shabad 1975), and omits higher order
contributions (e.g. see Melrose and Parle 1983a,b) to the vacuum 
polarization (for example, those that couple to photon absorption via
splitting) that become prominent in supercritical fields.  Furthermore,
plasma dispersion effects may be quite pertinent to soft gamma repeater
models (e.g. Bulik and Miller 1996), rendering them distinctly different
from pulsar scenarios.

\subsection{Pair Production Rate}

One-photon pair production is a first-order QED process that is quite
familiar to pulsar theorists.  It is forbidden in field-free regions due
to the imposition of four-momentum conservation, but takes place in an
external magnetic field, which can absorb momentum perpendicular to \bf
B\rm . The rate (Toll 1952, Klepikov 1954) increases rapidly with
increasing photon energy and transverse magnetic field strength, becoming
significant for $\gamma$-rays above the threshold, $\omega =
2/\sin\theta_{\rm kB}$, and for fields approaching $B_{\rm cr}$. When
the photon energy is near threshold, there may be only a few
kinematically available pair states, and the rate will be resonant at
each pair state threshold, producing a sawtooth structure  (Daugherty \&
Harding 1983, hereafter DH83).  For photon energies well above
threshold, the number of pair states becomes large, allowing the use of
a more convenient asymptotic expression for the polarization dependent
attenuation coefficient (Klepikov 1954, Tsai \& Erber 1974):
\begin{equation}
   T^{\rm pp}_{\parallel,\perp} = {1\over 2}{\alpha\over \lambar} B'
   \sin\theta_{\rm kB}\Lambda_{\parallel,\perp}(\chi),
  \label{eq:ppasymp}
\end{equation}

\begin{equation} 
   \Lambda_{\parallel, \perp}(\chi) \approx \left\{
     \begin{array}{lr} 
     (0.31, 0.15)\, \exp \mbox{\Large $(-{4\over 3\chi})$} & \chi \ll 1 
     \\ \\
     (0.72, 0.48) \, \chi^{-1/3} & \chi \gg 1
     \end{array} \right.          \label{eq:ppratlim}
\end{equation}
where $\chi \equiv (\omega/2)B'\sin\theta_{\rm kB}$.

In polar cap pulsar models (e.g. Sturrock 1971, Ruderman and Sutherland 1975), 
high energy radiation is usually emitted at very small angles
to the magnetic field, well below pair threshold (see Harding 1995, for
review).  The $\gamma$-ray photons will convert into pairs only after they have
traveled a distance $s$ comparable to the field line radius of curvature
$\rho$, so that $\sin\theta_{\rm kB} \sim s/\rho$.  From the above expression,
the pair production rate will be vanishingly small until the argument of the
exponential approaches unity, i.e. when $\omega B'\sin\theta_{\rm kB} \gtrsim
0.2$.  Consequently, pair production will  occur well above threshold when $B
\ll 0.1 B_{\rm cr}$ and the asymptotic expression will be valid.  However when
$B \gtrsim  0.1 B_{\rm cr}$, pair production will occur at or near threshold,
where the asymptotic expression has been shown to fall orders of magnitude
below the exact rate (DH83). In the present calculation, we approximate the
near-threshold reduction in the  asymptotic pair production attenuation
coefficient  by making the substitution, $\chi \rightarrow \chi/F$, where  $F =
1 + 0.42(\omega\sin\theta_{\rm kB}/2)^{-2.7}$ in equation~(\ref{eq:ppratlim}) 
(DH83).  Baring (1988) has derived an analytic expression for the one-photon
pair production rate near threshold which gives a result that agrees numerically
with the approximation of DH83.  

Yet even the near-threshold correction to the asymptotic rate becomes poor when
$B \gg  0.1 B_{\rm cr}$ and the photons with parallel and perpendicular
polarization produce pairs only (DH83) in the ground (0,0) and first excited
(0,1) and (1,0) states respectively.  Here \teq{(j,k)} denotes the Landau
level quantum numbers of the produced pairs.  Therefore when the local $B>0.1
B_{\rm cr}$, instead of the asymptotic form in equation~(\ref{eq:ppratlim}), we
use the exact, polarization-dependent, pair production attenuation
coefficient (DH83), including only the (0,0) pair state for $\parallel$
polarization:
\begin{equation}
   T^{\rm pp}_{\parallel} = \dover{\alpha\sin\theta_{\rm kB}}{\lambar \xi 
   |p_{\hbox{\sevenrm 00}}|}\,\exp(-\xi), \quad 
   \omega \ge 2/\sin\theta_{\rm kB}\quad ,
  \label{eq:tpppar}
\end{equation}
and only the sum of the (0,1) and (1,0) states for $\perp$ polarization:
\begin{equation}
   T^{\rm pp}_{\perp} = \dover{\alpha\sin\theta_{\rm kB}}{\lambar\xi 
   |p_{\hbox{\sevenrm 01}}|}\, (E_0 E_1 + 1 + p_{01}^2)\,\exp(-\xi), \quad
   \omega \ge \dover{1+(1 + 2B')^{1/2}}{\sin\theta_{\rm kB}}
  \label{eq:tppperp}
\end{equation}
where 
\begin{displaymath}
   E_0  = (1 + p_{01}^2)^{1/2}\quad , \quad  
   E_1  = (1 + p_{01}^2 + 2B')^{1/2}
\end{displaymath}
for
\begin{displaymath}
   |p_{jk}|  = \left[ \dover{\omega^2}{4} \sin^2\theta_{\rm kB} - 1 - (j+k)B' + 
   \left( \dover{(j-k)B'}{\omega\sin\theta_{\rm kB}} \right)^2\right]^{1/2}  
\end{displaymath}
and
\begin{equation}
   \xi = \dover{\omega^2}{2B'}\sin^2\theta_{\rm kB}\quad .    \label{eq:xi}
\end{equation}
Actually, both the asymptotic and exact mean-free paths ($1/T_{\rm pp}$) are
so small in fields where photons pair produce at threshold that it is, in fact,
not important which rate is used at very high field strengths (i.e. $B' \gtrsim
1$).  The pair production rate in this regime thus behaves like a wall at 
threshold and photons will pair produce as soon as they satisfy the kinematic
restrictions on \teq{\omega} given in equations~(\ref{eq:tpppar})
and~(\ref{eq:tppperp}).  The creation of bound pairs rather than free pairs
is possible in fields $B' \gtrsim 0.1$ (Usov \& Melrose 1995), but this should 
not affect the present calculation since we do not model the full pair cascade.

\subsection{Attenuation Lengths}

To assess the relative importance of photon splitting compared to pair
production through a dipole magnetic field, we compute the attenuation length
\teq{L}, defined to be the path length over which the optical depth is unity:
\begin{equation} 
   \tau(\theta, \erg)\; =\;\int_0^L T(\theta_{\rm kB},\, \omega )\, 
   ds\; =\; 1\quad ,        \label{eq:tau}
\end{equation}
where \teq{ds} is the pathlength differential along the photon momentum
vector \teq{{\bf k}}and $T$ is the attenuation coefficient for either
splitting, $T_{\rm sp}$, or pair production, $T_{\rm pp}$.  In this
paper, attenuation lengths are computed as averages over  polarizations
of the initial photon and, for splitting, sums over the final
polarization states.  Here \teq{\theta_{\rm kB}} and the photon energy
\teq{k^{\hat o} = \omega}  are functions of the position (e.g. see
equation~[\ref{eq:4momgr}]), specifically measured in the local inertial
frame, while \teq{\theta} is the colatitude of emission and \teq{\erg}
is the photon energy to an observer at infinity; our treatment of curved
spacetime is discussed immediately below.  In regions where the path
length is much shorter than both the scale length of the field strength
variation or the radius of curvature of the field, \teq{L} reduces to
the inverse of the attenuation coefficient.  In the calculation of the
splitting attenuation lengths, all three CP-permitted modes are assumed to operate.  The attenuation length behavior of the individual modes are similar.

\placefigure{fig:geometry}

We assume that test photons are emitted at the neutron star surface and
propagate outward, initially parallel or at a specified angle,
$\theta_{\rm kB,0}$ to the dipole magnetic field (see
Fig.~\ref{fig:geometry} for a depiction of the geometry).  Photon
emission in polar cap models of gamma-ray pulsars can occur above the
stellar surface (but see the discussion in Section~5), which would
generate attenuation lengths somewhat longer than those determined here,
due to the \teq{r^{-3}} decay of the field.  A surface origin of the
photons is chosen in this paper to provide a simple and concise
presentation of the attenuation properties.  We have included the
general relativistic effects of curved spacetime in a Schwarzschild
metric, following the treatment of Gonthier \& Harding (1994, GH94) who
studied the effects of general relativity on photon attenuation via
magnetic pair production.  GH94 included the curved spacetime photon
trajectories,  the magnetic dipole field in a Schwarzschild  metric and
the gravitational  redshift of the photon energy.  One improvement we
have made here to the  treatment of GH94 is to explicitly keep track of
the gravitational redshift of the photon energy as a function of
distance from the neutron star (see Appendix for details).  Our analysis
is confined to the Schwarzschild metric because the dynamical timescales
for gamma-ray pulsars are considerably shorter than their period (e.g.
\teq{P=0.15}sec. for PSR1509-58), so that rotation effects in the Kerr
metric can be neglected.  We have taken a neutron star mass, $M =
1.4\,M_\odot$ and radius, $R = 10^6$ cm in these calculations.

Fig.~\ref{fig:attenl} illustrates how the attenuation lengths for photon
splitting and pair production vary with energy for different magnetic
colatitudes of the emission point, for surface fields of
\teq{B_0=0.1B_{\rm cr}}, and \teq{B_0=0.7B_{\rm cr}}.  A field of $B_0 =
0.7B_{\rm cr}$ is the value of the polar surface  field derived from the
magnetic dipole spin-down energy loss (Shapiro \& Teukolsky 1983), using
the measured $P$ and $\dot P$ for PSR1509-58.  As noted by Usov \&
Melrose (1995), this is exactly twice the value of the surface field
given by formulae in other sources (Manchester  \& Taylor 1977, Michel
1991), which assume (inaccurately) that the  dipole magnetic moment
$\mu = B_0\,R^3$ rather than $\mu = B_0\,R^3/2$ for a uniformly
magnetized sphere of radius $R$. The other $\gamma$-ray pulsars have
surface field strengths in the range $1 - 9\times 10^{12}$ G, or 
$0.02 - 0.2\,B_{\rm cr}$ (the Crab and Vela pulsars have fields around
$0.2\,B_{\rm cr}$).  Note that the attenuation lengths in
Fig.~\ref{fig:attenl} are for unpolarized radiation; the curves for
$\parallel$ and $\perp$ polarization states look very similar. 

\placefigure{fig:attenl}

The curves in Fig.~\ref{fig:attenl} have a power-law behavior at high
energies, i.e. for attenuation lengths much less than $10^{6}$ cm, where
the dipole field is almost uniform in direction and of roughly constant
strength.  They also exhibit sharp increases at the low energy end,
where photons begin to escape the magnetosphere without attenuation.  We
may estimate the behavior of the power-law portions of the attenuation
length curves in Fig~\ref{fig:attenl} as follows. Since the photons are
assumed to initially propagate parallel to the field, the field
curvature will give propagation oblique to the field only after
significant distances are traversed, so that the obliquity of the photon
to the field scales, to first order, as the distance travelled, 
\teq{\sin\theta_{\rm kB}\propto s}.  Inserting this in 
equation~(\ref{eq:splitotrate}) gives a photon splitting attenuation
coefficient  \teq{\propto s^6} i.e. an optical depth \teq{\propto \erg^5
s^7}, since \teq{T_{\rm sp}\propto\erg^5}. Inversion then indicates that
the attenuation length should vary as \teq{L \propto \erg^{-5/7}}: this
is borne out in Fig.~\ref{fig:attenl}. For $B_0 \gtrsim 0.1B_{\rm cr}$,
pair production occurs as soon as the threshold $\erg_{\rm th} =
2/\sin\theta_{\rm kB}$ is crossed (cf. Section 2.2) during the photon
propagation in the magnetosphere.  Essentially, due to the enormous
creation rate immediately above the threshold, this energy serves as an
impenetrable ``wall'' to the photon.  Again, since  \teq{\sin\theta_{\rm
kB} \propto s} in the early stages of propagation, the pair production
attenuation length should scale as \teq{L\propto 2/\erg}.  These
proportionalities hold in both curved and flat spacetime since  general
relativistic effects distort spacetime in a smooth and differentiable
manner (see the Appendix).  However, the attenuation lengths computed in
the Schwarzschild metric are about a factor of 1.5 lower than those
computed in flat spacetime (Baring \& Harding 1995b). 

The photon splitting attenuation coefficient we have used is strictly
valid only below pair threshold.  Hence, the attenuation lengths for
splitting depicted in Fig.~\ref{fig:attenl} can be regarded as only
being symbolic when they exceed those for pair production, since then
pair threshold is reached before splitting occurs.  No technically
amenable general expressions for the rate of splitting above pair threshold
exist in the physics literature.  But the vicinity of parameter space
just below pair threshold is the regime of  importance for $\gamma$-ray
pulsar models, where the emitted photons propagate until they either
split or they reach pair threshold, in  which case they pair produce. 
The attenuation length curves near the crossover points in 
Fig.~\ref{fig:attenl} for $B_0 = 0.7B_{\rm cr}$ will require  inclusion
of high energy corrections to the attenuation coefficient  (Stoneham
1979) that arise as the $\gamma\to e^+e^-$ threshold is approached. 
Currently work is in progress to compute these modifications (Baring and
Harding 1996, in preparation), and preliminary results indicate that the
rate in equation~(\ref{eq:splitotrate}) is quite accurate for
\teq{B\lesssim 0.2B_{\rm cr}}, but increases by factors of at most a few
for \teq{B=0.7B_{\rm cr}} and \teq{\omega =2}, as mentioned in
Section~2.1 above.

\subsection{Escape Energies}

The energy at which the attenuation length becomes infinite defines the
\it escape energy\rm, below which the optical depth is always \teq{\ll
1}, and photons escape the magnetosphere; the existence of such an
escape energy is a consequence of the \teq{r^{-3}} decay of the dipole
field.  Escape energies of unpolarized photons for both photon splitting
and pair production are shown in Fig.~\ref{fig:escape} as a function of
magnetic colatitude $\theta$ of the photon emission point for different
values of magnetic field strength (see also Harding, Baring and
Gonthier, 1996).  The escape energies clearly decline with \teq{\theta}
and are monotonically decreasing functions of \teq{B} for the range of
fields shown.   The divergences as \teq{\theta\to 0} are due to the
divergence of the field line radius of curvature at the poles.  There
the maximum angle \teq{\theta_{\rm kB}} achieved before the field falls
off and inhibits attenuation is proportional to the colatitude
\teq{\theta}.  For photon splitting, since the rate in
equation~(\ref{eq:splitotrate}) is proportional to
\teq{\omega^5\sin^6\theta_{\rm kB}}, and therefore also the attenuation
length \teq{L}, it follows that the escape energy scales as 
\teq{\erg_{\rm esc} \propto \theta^{-6/5}} near the poles (see also 
Fig.~\ref{fig:icescape}) as is determined by the condition \teq{L\sim
R}.  For pair production, the behaviour of the rate (and therefore
\teq{L}) is dominated by the exponential form in
equation~(\ref{eq:ppratlim}), which then quickly yields a dependence
\teq{\erg_{\rm esc} \propto \theta^{-1}} near the poles for
\teq{B_0\lesssim 0.1B_{\rm cr}}.  This behaviour extends to higher
surface fields because production then is at threshold, which determines 
\teq{\erg_{\rm esc}\sim 2/\theta_{\rm kB} \propto \theta^{-1}}. At high
fields, $B_0 \gtrsim 0.3B_{\rm cr}$, there is a saturation of the photon
splitting attenuation lengths and escape energies, due to the
diminishing dependence of $B$ in the attenuation coefficient.  Likewise,
there is a saturation of the pair production escape energy at fields
above which pair production occurs at threshold. The pair production
escape energy curves are bounded below by the  pair threshold
$2/\sin\theta_{\rm kB}$ and merge for high $\theta$, at the pair rest
mass limit, $\erg = 2$, blueshifted by the factor $(1-2GM/Rc^2)^{-1/2} \sim
1.3$.  Note that photon splitting can attenuate photons well below pair
threshold. For low fields, pair production escape energies are below
those for splitting, but in high fields, splitting escape energies are
lower at all $\theta$.  The escape energies are roughly equal for $B_0
\sim 0.3 B_{\rm cr}$.  

\placefigure{fig:escape}

The effects of curved spacetime are quite significant when compared to the
attenuation lengths and the escape  energies obtained assuming flat spacetime. 
A comparison of the escape energies for splitting and pair production, computed
in flat and curved spacetime, is shown in Fig.~\ref{fig:escurvflat}. 
The largest effects are due
to the increase of the surface dipole field strength by roughly a factor of
1.4, and the correction for the  gravitational redshift of the photon, which
increases the photon energy by roughly a factor 1.2 in the local inertial frame
at the neutron star surface compared to the energy measured by the observer in
flat space (see the Appendix).  The combination of these effects decreases
the photon splitting escape energy by  a factor of about 2 compared to flat
spacetime.  The decrease in escape energy for pair production is also a factor
of about 2, except at the largest values of $\theta$ and $B'$, 
where the pair rest mass limit is reached (cf. Fig.~\ref{fig:escape}) 
The escape energy is then no longer dependent on
field strength, and the ratio of the curved to flat space escape energy 
is just the redshift of the photon energy ($\sim 0.8$) 
from the conversion point.  This is achieved in the upper right hand corner
of the figure; photon splitting has no such strict limit.
The ratios also become insensitive to \teq{\theta} near the poles since there
the photons move almost radially, thus traveling along straight trajectories,
and the curved-space correction to the field is not changing rapidly with
colatitude.  The curvature of the photon trajectory in a Schwarzschild 
metric does not affect the escape energies, to first order, except in the case
of emission at large colatitudes, where the photon wavevector makes a large
angle to the radial direction.  

\placefigure{fig:escurvflat}

High energy emission from curvature radiation, inverse Compton or
synchrotron by relativistic particles with Lorentz factor $\Gamma$  will
not beam the photons precisely along the magnetic field, but within some
angle $\sim 1/\Gamma$ to the field. Fig.~\ref{fig:icescape}  illustrates
the effect on the escape energies of a non-zero angle of emission of the
photons, for the case where the photons are emitted at angles toward the
dipole axis. We have chosen the angle  $\theta_{\rm kB,0} = .01$ rad $(= 
0.57^\circ$) because it would be the angle at which photons with $\epsilon
\sim 100$ would be emitted through the cyclotron upscattering process,
 $\theta_{\rm kB,0} \simeq B'/\epsilon$ (Dermer 1990). 
For emission angles $\theta_{\rm kB,0} = 0$ in
Fig.~\ref{fig:icescape}a, which plots \teq{\erg_{\rm esc}} for photon
splitting, $\erg_{\rm esc} \propto B_0^{-6/5}$ for $B_0 \lesssim
0.3B_{\rm cr}$ and $\erg_{\rm esc} \propto \theta^{-6/5}$,  for $\theta
\lesssim 20^\circ$, dependences that naturally follow  from the form of
equation~(\ref{eq:splitotrate}). Generally, the escape energy is
insensitive to the emission angle for $\theta \gtrsim 10 \theta_{\rm
kB,0}$.  For small angles, the escape energy decreases and the 
$\theta_{\rm kB,0} = .57^\circ$ curves
flatten below the  $\theta_{\rm kB,0} = 0$ curves, converging as
\teq{\theta\to 0} to an energy that is proportional to
\teq{(B_0\sin\theta_{\rm kB,0})^{-6/5}} (see Eq.~[\ref{eq:ergs}]).  This
convergence is a consequence of the field along photon trajectories that
originate near the pole being almost uniform and tilted at about angle
\teq{\theta_{\rm kB,0}} to the photon path.  In
Fig.~\ref{fig:icescape}b, the same effect is seen for pair creation, but
this time the ``saturation'' is at the redshifted threshold energy 
$2(1-2GM/Rc^2)^{1/2}/\sin\theta_{\rm kB,0}$, and is independent of
\teq{B_0}.  We note that this behaviour at low colatitudes was observed,
in the case of pair creation in flat spacetime, by Chang, Chen and Ho (1996). 

\placefigure{fig:icescape}

An obvious exception to this expected behaviour is seen in
Fig.~\ref{fig:icescape}b for the  $B' = 3.1$ curve, where the escape
energy is actually \it larger \rm at small colatitudes 
(\teq{1^\circ\lesssim\theta\lesssim 10^\circ}) when the emission angle
$\theta_{\rm kB,0}$ is increased. This counter-intuitive result can be
understood with the aid of Fig.~\ref{fig:wsinthet}, which shows the
increase in $\sin\theta_{\rm kB}$,  and $\omega\sin\theta_{\rm kB}$,
determined in the local inertial frame, with path length \teq{s} along the
photon trajectory. Note that (i) the \teq{\theta_{kB,0}=0} curves increase in
proportion to \teq{s} when \teq{s/R\ll 1}, as described in the Appendix,
and (ii)  the $\sin\theta_{\rm kB}$ curves increase logarithmically with 
$s/R$ when \teq{s/R} is not very small.  In this large field, pair
production occurs when the threshold $\omega\sin\theta_{\rm kB} = 2$ is
crossed, at the same path length for both $\theta_{\rm kB,0} = 0$ and
$\theta_{\rm kB,0}  = 0.57^\circ$. The differences in the
photon trajectories (which are almost radial) for these two cases are so
small that \teq{s} effectively represents the same height above the stellar
surface for both
\teq{\theta_{\rm kB,0}}.  Since \teq{\omega\sin\theta_{\rm kB}\approx 2}
defines the pair creation ``wall''  for both photon paths, the only
difference in escape energies  is due to the factor of
\teq{\sin\theta_{\rm kB}} at the front of the  pair creation rates in
equations~(\ref{eq:ppasymp})--(\ref{eq:tppperp}). Hence, at the point of
pair creation, the value of $\sin\theta_{\rm kB}$  is smaller for the
$\theta_{\rm kB,0} = 0.57^\circ$ case, and therefore the escape energy
is larger. In flat spacetime, which is not depicted in
Figs.~\ref{fig:icescape} or~\ref{fig:wsinthet}, the crossover point of
the $\sin\theta_{\rm kB}$ curves occurs at the same $s/R$ value as pair
threshold, so that the escape energies are the same at this colatitude
for the two cases (this situation was also observed by Chang, Chen and Ho
1996).   Note that as photon splitting does not have the same sudden
onset as pair creation, it takes place over a range of path lengths, mostly
around \teq{0.1\lesssim s/R\lesssim 2}.  Over this range,
\teq{\sin\theta_{\rm kB}} in Fig.~\ref{fig:wsinthet} is generally larger
for the \teq{\theta_{\rm kB,0}=0.57^\circ} case so that the splitting
escape energy is correspondingly shorter than for emission parallel to
the field, as is evident in Fig.~\ref{fig:icescape}a.

\placefigure{fig:wsinthet}

\section{CASCADE SPECTRA}

Here we describe briefly our Monte Carlo simulation of photon propagation
and attenuation via splitting and pair creation in neutron star magnetospheres,
together with results for single (Section 3.2) and multiple (Section 3.3)
generations of photon splitting.

\subsection{Monte Carlo Calculation}

We model the spectrum of escaping photons from a cascade above a neutron star
polar cap, including both photon splitting and pair production, by means of a
Monte Carlo  simulation.  The free parameters specified at the start of the
calculation are the magnetic colatitude $\theta$, the angles $\theta_k$ and
$\phi_k$ (see Fig.~\ref{fig:geometry}), the spectrum, 
the height above the surface $z_0 = r-R$
of the photon emission, and the surface magnetic field strength $B_0$ (note
that entities with subscripts `0' designate determination at the stellar
surface).  From these quantities, and assuming that $\phi = 0$ without loss of
generality, we compute the four-vectors of the photon position and momentum
that are carried through the computation. Injected photons are sampled from a
power-law distribution,
\begin{equation} 
   N(\erg) = N_0\erg^{-\alpha},  \quad  \erg_{min} < \erg < \erg_{max}
  \label{eq:N}
\end{equation}
Polarization is chosen randomly to simulate unpolarized emission; this can
be altered, as desired, for any postulated emission mechanism.

The path of each input photon is traced through the magnetic field, in
curved spacetime, accumulating the survival probabilities for splitting,
$P_{\rm surv}^s$, and for pair production, $P_{\rm surv}^p$, independently:
\begin{equation}
   P_{\rm surv}(s) = \exp\Bigl\{-\tau(s)\Bigr\} 
\end{equation}
where
\begin{equation}
   \tau(s) = \int_0^s T(\theta_{\rm kB}, \omega ) ds'
\end{equation}
is the optical depth along the path.  These survival probabilities
implicitly depend on the origin \teq{{\bf r}_0} of the photon and its
energy \teq{\erg} at infinity.  In computing the attenuation lengths
(Section 2.3), the photon was assumed to split when the survival
probability reaches $1/e$, i.e. when equation~(\ref{eq:tau}) is
satisfied. In the cascade simulation, the photon may either split or
pair produce. The fate of each cascade photon is determined as follows:
if the combined survival probability, $P_{\rm surv}^sP_{\rm surv}^p >
\Re_1$, where $\Re_1$ is a random number between 0 and 1, 
chosen at the emission point,
then the photon escapes;  if not, then if the probability that the
photon survives splitting but not pair production, $P_{\rm
surv}^s(1-P_{\rm surv}^p)/(1-P_{\rm surv}^s P_{\rm surv}^p) > \Re_2$,
where $\Re_2$ is a second random number, then the photon pair produces;
otherwise, the photon splits.  When the photon splits, the energy of one
of the final photons is sampled from the distribution given in 
equation~(\ref{eq:splitdiffrate}) and their polarizations are chosen
from the  branching ratios given in equation~(\ref{eq:splitpolrate}).
The  energy of the second photon from the splitting is determined simply
from energy conservation, since both final photons are assumed to be
collinear in the direction of the parent photon. Each final photon is
then followed in the same way as the injected photon, with a call to a
recursive procedure that stores photon energies and positions through
many generations of splitting.  When the photon pair produces, the code
does not follow the radiation from the pairs but simply returns to the
previous cascade generation.  For field strengths typical of gamma-ray
pulsars, the pair radiation, most probably synchrotron or inverse
Compton, will not contribute significantly  at the energies near the
escape energy for the cascades where all splitting modes operate.  An
exception to this may occur for supercritical surface fields, where
synchrotron photons acquire most of the energy of their primary
electrons.  When all splitting modes operate, the number of pair
production events is a small fraction of the number of splitting events
for $B_0 = 0.7B_{\rm cr}$.  The cascade photons are followed through
many generations of splitting until all of the photons either escape
or pair produce.  The escaping photons are binned in energy and
polarization.

\subsection{Partial Splitting Cascade}

For pulsar applications with subcritical fields,  as discussed in
Section 2.1, it is probable that the splitting modes allowed by CP
invariance are further limited by kinematic selection rules to only  the
$\perp \rightarrow \parallel\parallel$ mode.  This restriction may be
confined to regimes of weak vacuum dispersion and may also depend on 
subtleties such as field non-uniformity. Such selection rules would 
effectively prevent splitting cascades since $\perp$ photons could split
only into  $\parallel$ photons which do not split.  Here we compute the
emergent spectra in this type of cascade, a partial cascade, where
$\perp$ mode photons can either pair produce or split into $\parallel$
mode photons, while the $\parallel$ mode photons may only pair produce. 
There is a limit of two cascade generations: one splitting and one pair
production.  The input spectrum is a power-law (Eq.~[\ref{eq:N}]) with
the  parameters: $\erg_{min} = 10^{-3}$, $\erg_{max} = 100$ and $\alpha
= 1.6$. The value of the index $\alpha$ is chosen to match the power-law
fit of the OSSE spectrum of PSR1509-58 (Matz et al. 1994).  The maximum
energy of the input spectrum $\erg_{max} = 100$ is chosen to fall above
the 30  MeV maximum possible cutoff or turnover energy of the observed
PSR1509-58 spectrum.  
For these runs, injection of 5 - 10 million photons are required
to give adequate statistics. The number of pairs produced relative to 
photons in these
partial splitting cascades is obviously higher than in the full
cascades examined in the next section.  
Note that in more complete gamma-ray pulsar
models that include the pair radiation, multiple generations of splitting 
might still be possible, being interspersed with generations of conventional 
synchrotron/pair cascading.

\placefigure{fig:specpar}

Figure~\ref{fig:specpar} shows partial splitting cascade spectra in each
final polarization mode for photons injected parallel to the local
magnetic field at different magnetic colatitudes.  The spectra for the
two polarization modes are cutoff at slightly different energies,
reflecting the different escape energies for splitting, which cuts off
the $\perp$ mode, and for pair production, which cuts off the
$\parallel$ mode.  There is a slight bump below the cutoff in the $\parallel$ 
mode spectrum, due to escaping photons from $\perp \rightarrow
\parallel \parallel$ mode splitting, but only an attenuation cutoff in the
$perp$ mode spectrum.  
Figure~\ref{fig:icspecpar} illustrates the effect of injecting photons
at an angle (in this case $\theta_{\rm kB,0} = 0.57^\circ = 0.01$ radians) 
to the local magnetic field direction, toward the magnetic dipole axis. 
The high-energy cutoff decreases, compared to the case of injection
parallel to the field, only in the $\perp$ mode spectrum and 
not at all in
the $\parallel$ mode spectrum. This behavior is due to the existence of
a threshold for pair production, but not for splitting and can be seen
from  Figs.~\ref{fig:icescape}a and~\ref{fig:icescape}b.  For field
strengths well above $B' = 0.1$, where photons pair produce at
threshold, the pair escape energy is much less sensitive to increases in
$\theta_{\rm kB,0}$ than is the splitting escape energy.  The partial
cascade spectra therefore  become more highly polarized at small
colatitudes when $\theta_{\rm kB,0}$ is increased.

\placefigure{fig:icspecpar}

This effect of strong polarization, both in the energy of the
spectral cutoffs and the spectral shape just below the cutoffs,
all but disappears when photon splitting is omitted from the
calculation, thereby defining a characteristic signature of the action
of \teq{\split}.  Pair production has much less distinctive polarization
features.  For example, from  equations~(\ref{eq:tpppar})
and~(\ref{eq:tppperp}),  the ratio of the cutoff energies at pair
creation threshold between the polarization states is
\teq{(1+\sqrt{1+2B'})/2}.  For surface fields of \teq{B'=0.7}, threshold
is crossed during photon propagation in regions with much lower fields,
typically \teq{B'\sim 0.1}, so that the spectral cutoff (or escape
energy) differs only by about 5\% between polarizations; such a
difference would be virtually invisible in the emission spectra. Clearly
then, splitting is primarily responsible for polarization features
shown.

\subsection{Full Splitting Cascade}

We now present model cascade spectra for the case where all three
photon splitting modes allowed by CP invariance, $\perp \rightarrow
\parallel \parallel$, $\perp \rightarrow \perp\perp$ and $\parallel
\rightarrow  \perp\parallel$, are operating, and multiple generations of
splitting can occur.  These cascades also allow for pair production by
photons of either mode.  As noted above, for the field strength of $B_0'
= 0.7$ used in the spectral models for PSR1509-58, pair production
occurs in less than $10\%$ of conversions.  

Figure~\ref{fig:specpol} shows full splitting cascade spectra in each
final polarization mode for 2 million photons injected parallel to the
local magnetic field (in curved spacetime) at different magnetic
colatitudes.  Each cascade spectrum shows a cutoff at roughly
the splitting escape energy for that colatitude (compare to
Fig.~\ref{fig:escape}), and a bump below the cutoff from the escaping
cascade photons.  The size of the bump is a function of the number of
photons attenuated above the cutoff, which is dependent on the ratio of
the maximum input energy, $\erg_{max}$, and the escape energy.  For
these models, the size of the cascade bump grows with increasing
$\theta$ because $\erg_{max}$ is held constant while the escape energy
is decreasing.  The number of splitting generations ranges from 12 when
$\theta = 30^\circ$ to  3 when $\theta = 2^\circ$.  The size of the
cascade bump at a particular $\theta$ could of course be larger or
smaller if $\erg_{max}$ were increased or decreased, but the positions
of the cutoffs would not vary.  The spectrum of the bump is polarized,
with a well-defined zero in polarization that is a characteristic
signature of the splitting cascade (see Baring 1995).  Note that
the polarization  modes have reversed their flux dominance in the
cascade bump compared to the partial splitting cascade case.

\placefigure{fig:specpol}

Although we have injected unpolarized photons in these calculations for
simplicity, the relative flux (i.e. spectra integrated over energies)
of the two polarization modes generally has a complicated dependence
on the branching ratios for splitting defined by
equation~(\ref{eq:splitpolrate}), due to the cascading process and the
non-uniformity of the field.  Notwithstanding, the polarization at a
given energy does not exceed a limiting value of 3/7 
(Baring 1991). The cascade spectra for injection of
polarized photons resemble the spectra in Fig.~\ref{fig:specpol},
though deviations from Fig.~\ref{fig:specpol} are not exactly
proportional to the degree of polarization of the injection spectrum due
to the inherent complexity of the interplay of polarization states in
the cascade. 

\placefigure{fig:icspecpol}

As shown in
Figure~\ref{fig:icspecpol}, injecting photons at an angle to the local
field again has a much larger effect at small colatitudes (i.e. for
$\theta \lesssim 100\theta_{\rm kB,0}$). The
high-energy cutoffs in both modes now decrease in energy compared 
to the case of injection
parallel to the field, and the sizes of the cascade bumps are larger,
both being consequences of decrease in escape energy (see
Fig.~\ref{fig:icescape}a).  This effect is larger at smaller
colatitudes.

\section{PHOTON SPLITTING CASCADE MODELS FOR PSR1509-58}

The multiwavelength spectrum of PSR1509-58, compiled from radio to TeV
energies (Thompson 1996), shows that the peak in the power output from
this pulsar, as is the case for most other $\gamma$-ray pulsars, falls
in the $\gamma$-ray band.  
Figures~\ref{fig:datpar}--\ref{fig:daticunpol} show the high energy
portion of this spectrum, near the cutoff, which we compare with our
model  spectra at different emission colatitudes.  No formal procedure
for fitting the data with the model was followed, since a simple visual
comparison demonstrating the spectral cutoff is sufficient for the
scientific goals of this paper. The $\epsilon^2\,F(\epsilon)$ format 
plots equal
energy per logarithmic decade and clearly demonstrates the need for a
cutoff or sharp turnover somewhere between the highest OSSE detected
point at 3 MeV and the lowest EGRET upper limit at 30 MeV.  Although
there appears to be a discontinuity between the GINGA data points below
100 keV and the OSSE data points, it is common for separate fits of data
from two different detectors to produce disparate results, even in the
same energy range. Furthermore, the $\epsilon^2\,F(\epsilon)$ format 
tends to magnify
the differences.  The difference in spectral index of the Ginga and OSSE
fits probably indicates a true break in the power-law spectrum around
100 keV. We have taken the OSSE index for the input spectrum for our
cascade simulation since it most accurately measures the observed
spectrum at the energies of importance for the model.  The offset between
the Ginga and OSSE data (or their different spectral indices) does not
impact the  conclusions of this paper, since the cascade formation is
determined by the photon population in the upper end of the OSSE range.  
Note that while EGRET has obtained upper limits to the pulsed emission
above around 30 MeV, there are earlier reports of a marginal detection
by COS-B (e.g.. Hartmann et al. 1993), with data points lying above the
EGRET limits.  This apparent discrepancy remains to be resolved, and we
opt here to consider only the later and superior EGRET observations.
The Comptel point and limits in Figs.~\ref{fig:datpar}--\ref{fig:daticunpol}
are a preliminary analysis of data from VP23 (Hermsen et al. 1996), 
showing pulsed flux at 0.75 - 1 MeV and upper limits for the pulsed interval 
(50\%) of the light curve.

The cutoffs in the model photon splitting cascade spectra in
Figures~\ref{fig:datpar}--\ref{fig:daticunpol}
do in fact fall in the energy range 3--30
MeV for colatitudes less than around $30^\circ$.  At colatitudes greater
than $\sim 30^\circ$ the cutoffs are lower and are in severe conflict
with the OSSE data points.   The standard polar cap half-angle in flat
spacetime, $\sin\theta = (\Omega R/c)^{1/2}$, for  PSR1509-58 is
$2.14^\circ$.  Although curved spacetime corrections to the magnetic
dipole field tend to very slightly decrease the polar cap size (GH94),
the polar cap may be larger than the standard size due to distortion of
the field near the light cylinder by plasma loading (Michel 1991).
The results presented here assume, for simplicity, a single colatitude
of emission for each, i.e. a polar rim rather than an extended cap.
It is easy to envisage that a range of polar cap emission locations will
produce a convolution of the spectra presented here, thereby generating
a spectral turnover corresponding to the maximum colatitude of the cap,
with steeper emission extending up to a cutoff defined by the minimum
colatitude.  The EGRET upper limits cannot really discern between a
sharp cutoff or a more modest turnover above the Comptel energy range
and so cast little light on the emission as a function of colatitude
when \teq{\theta\lesssim 2^\circ}. 

\placefigure{fig:datpar}
\placefigure{fig:daticpar}

The partial splitting cascade spectra, shown in Figs.~\ref{fig:datpar}
and~\ref{fig:daticpar}, exhibit only modest cascade bumps just below the
cutoff.  The limits on colatitude of the model spectra are essentially
determined by the cutoff energy and are restricted by the lowest EGRET
upper limit to $2^\circ \lesssim  \theta
\lesssim 25^\circ$ in the $\theta_{\rm kB,0} = 0$ case, and $\theta
\lesssim 25^\circ$ in the  $\theta_{\rm kB,0} = 0.57^\circ$ case, where
no lower limit to the colatitude is imposed by the observations (see below). 
The model spectra for $\theta = 2^\circ$ and $5^\circ$ are only marginally
consistent with the upper limits.  The final revision of the Comptel
data for PSR1509-58 (Bennett et al., in preparation) 
may require raising the lower bounds to the 
colatitude of emission obtained in this model/data comparison.
The cutoff energies of these polarization-averaged spectra are somewhat
larger than the cutoff energies of the full cascade spectra  (see
Figs.~\ref{fig:datunpol} and~\ref{fig:daticunpol}) because the
$\parallel$ mode escape energies are determined solely by pair
production, whose escape energies generally exceed those of splitting at
this field strength (see Fig.~\ref{fig:escape}).  This is especially
pronounced in the $\theta_{\rm kB,0}=0.57^\circ$ case, due to the fact
that the pair production escape energy is insensitive to the photon
emission angle for $B \gg 0.1$, as is illustrated in
Fig.~\ref{fig:icescape}b.

The full cascade spectra, shown in Figs.~\ref{fig:datunpol}
and~\ref{fig:daticunpol}, have distinctive bumps  below the cutoff due
to the redistribution of photon energies via splitting.   The size of
the cascade bump further limits the magnetic colatitudes to $\theta \lesssim
5^\circ$ to avoid conflict with the Comptel upper limits. The lowest EGRET
upper limit restricts the colatitudes to $5^\circ \gtrsim \theta 
\gtrsim 2^\circ$ in the
case of emission parallel to the field  (Fig.~\ref{fig:datunpol}).   In
the case of emission at angle $\theta_{\rm kB,0} = 0.57^\circ$ 
(Fig.~\ref{fig:daticunpol}),  the cutoff energy in the cascade spectra
saturates at small $\theta$ at an energy of 25 MeV (see 
Fig.~\ref{fig:icescape}a).  Consequently there is no low-energy limit to
$\theta$ in this case.  For larger values of $\theta_{\rm kB,0}$, the
spectral cutoffs would saturate at larger values of $\theta$ and at
lower energies.  We can estimate the dependence of this saturation
escape energy, $\erg_{\rm esc}^{\rm sat}$, on $\theta_{\rm kB,0}$ and
$B'$ using the expression for the splitting attenuation coefficient in
equation~(\ref{eq:splitotrate}).  Assuming that $1/T_{\rm sp} \simeq R$
approximately gives the escape energy:
\begin{equation} 
   \erg_{\rm esc}^{\rm sat} \simeq 0.077\,(B'\sin\theta_{\rm kB,0})^{-6/5}
   \;\; ,\quad B' \lesssim 0.3.    \label{eq:ergs}
\end{equation}
This formula quite accurately reproduces the escape energies in
Figs.~\ref{fig:datunpol} and~\ref{fig:daticunpol} since they are only
weakly dependent on \teq{R}, specifically 
\teq{ \erg_{\rm esc}^{\rm sat}\propto R^{-1/5}}.
When $\erg_{\rm esc}^{\rm sat} \le 7.8$ (i.e. 4 MeV), cascade spectra at
all colatitudes cutoff below the lowest possible observed  cutoff
energy for the PSR1509-58 spectrum.  From equation~(\ref{eq:ergs}), 
this occurs, for surface emission, at $\theta_{\rm kB,0} \gtrsim 0.03$
for $B_0' = 0.7$.  Therefore, splitting cascade spectra from photons
emitted at larger angles to the field will not be compatible with the
spectrum of PSR1509-58.  For emission at some distance above the
surface, the limit on $\theta_{\rm kB,0}$ would be higher since it
depends inversely on local field strength.

\placefigure{fig:datunpol}
\placefigure{fig:daticunpol}

All the model spectra in Figures~\ref{fig:datpar}--\ref{fig:daticunpol}
assume emission at the neutron star surface.  Emission above the surface
would produce higher cutoff energies at a given colatitude, due to the
decrease in the dipole field strength with $r$. The upper limits on 
colatitude
stated above would therefore be less restrictive for non-surface
emission.  Furthermore, when the field strength at the emission point is
$B \sim 0.3 B_{\rm cr}$ (at height $30\%$ of the neutron star radius)
the splitting and pair production escape energies are comparable,
reducing the size of the splitting cascade bumps in all cases.  At
higher altitudes above the surface, pair production dominates the photon
attenuation and conventional pair cascades (e.g. Daugherty \& Harding
1996) would operate.  Synchrotron radiation from the pairs would then
result in a significantly softer emergent spectrum than the input
power-law above the cyclotron energy (\teq{\sqrt{1+2B'}-1\approx
280}keV at the stellar surface, lower at greater radii).  Consequently, 
in order to match the observations, the input
power-law would have to be harder, and because of the remoteness of the
emission point from the stellar surface, the colatitude \teq{\theta} of
emission would have to be increased substantially.

\section{DISCUSSION}

The results of this paper demonstrate that magnetic photon splitting can
have a significant effect on $\gamma$-ray emission from the higher field
($B_0 \gtrsim 10^{13}$ G) pulsars.  It can attenuate the $\gamma$-ray 
spectrum at lower energies than magnetic pair production and will do so
without the creation of electron-positron pairs.   We have found that in
low fields ($B_0 \lesssim 0.3B_{\rm cr}$) and  $\theta_{\rm kB,0} = 0$
initially, photon splitting attenuation lengths are never shorter than
those for pair production.  In high fields ($ B_0 \gtrsim 0.3B_{\rm
cr}$), photon splitting lengths fall below those for pair production
below a certain energy which depends on the colatitude $\theta$.  Photon
splitting escape energies fall below pair production escape energies for
$B_0 \gtrsim 0.5\,B_{\rm cr}$, so that splitting may produce an observable
signature for $\gamma$-ray pulsars having strong magnetic fields: high
energy spectral cutoffs that are quite polarization-dependent.  While
pair creation alone will also generate such cutoffs, their dependence on
photon polarization is far diminished from when splitting is active. 

We have modeled the shape  of such spectral cutoffs through simulation of
photon splitting cascades near the neutron star surface for the case of
PSR1509-58.  Two types of cascades result from different assumptions about
the selection rules governing the photon splitting modes: the ``full 
splitting cascades" occur when three modes limited only by CP selection 
rules operate and the ``partial splitting cascades" occur when only one
mode permitted by kinematic selection rules operates.  
In the full cascades, splitting
dominates the attenuation while in the partial cascades, pair production
ultimately limits the rate at which photon energy degrades.  However, the
partial cascades show a distinct polarization signature due to the different
escape energies for splitting and for pair production. 
The resulting PSR1509-58 model spectral cutoffs due to splitting and pair
production fall in the required range for virtually all colatitudes 
$\lesssim 25^\circ$.  However, the shape of the spectrum of full splitting
cascades, due to the large reprocessing bump, is compatible with the 
data only for a very small range of colatitudes, $\theta \lesssim 5^\circ$.  
From these
results we conclude that, although photon splitting is capable of producing 
spectral cutoffs well below EGRET energies regardless of which selection rules 
govern the splitting modes, the partial splitting cascades have a much
larger range of phase space in which to operate.

Attenuation through magnetic pair production and photon splitting 
near the polar cap will produce $\gamma$-ray spectral cutoffs that should
be roughly a function of surface magnetic field strength, although other
parameters such as polar cap size will come into play.  Thus the 
$\gamma$-ray pulsar PSR0656+14, having the second highest surface field 
of $9.3 \times 10^{12}$ G, should have a cutoff energy between that of
PSR1509-58 and the other $\gamma$-ray pulsars.  In fact the unusually large
spectral index of 2.8 measured by EGRET (Ramanamurthy et al. 1996) may
be a pair production/photon splitting cutoff.
 
It is thus possible to  understand why PSR1509-58, with the highest
magnetic field of all the  $\gamma$-ray pulsars, has by far the lowest
spectral cutoff energy and is the only $\gamma$-ray pulsar not detected
by EGRET. In the case of Vela (Kanbach et al. 1994), Geminga
(Meyer-Hasselwander et  al. 1994) and 1055-52 (Fierro et al. 1993), the
spectral cutoffs observed by EGRET at a few GeV are consistent with
one-photon pair production cascades (Daugherty and Harding 1982, 1996). 
Although the escape energies at the neutron star surface for the
spin-down fields of these pulsars ($B_0 \sim 2 - 6 \times 10^{12}$ G) is
below 1 GeV (see Fig.~\ref{fig:icescape}a), curvature radiation from
primary electrons  at one to two stellar radii above the surface will
have pair production escape energies of several GeV.  However, when the
surface field exceeds $\sim 10^{13}$ G, photon splitting becomes the
dominant attenuation mechanism in the electromagnetic cascades. 
In addition, the primary electrons may lose energy to resonant Compton
scattering of thermal X-rays from the neutron star surface (Sturner
1995), rather than to curvature radiation, limiting their acceleration
to much lower energies, typically $\gamma \sim 100$. The resulting
upscattered $\gamma$-ray spectrum is radiated much closer to the surface
and will be cut off by photon splitting well below the EGRET energy range.
It is important to emphasize that pair creation acting alone suffices
to inhibit GeV emission in pulsars with spin-down fields as high as
PSR1509-58, and splitting significantly enhances the attenuation and
pushes spectral cutoffs to lower energies.

If resonant Compton scattering losses limit the polar cap particle 
acceleration energies to $\gamma \ll 10^6$ when $B \gtrsim 10^{13}$ G,
then the primary particles will radiate $\gamma$-rays via the cyclotron
upscattering process or CUSP (Dermer 1990).  CUSP radiation would then
provide the seed photons for the splitting cascade.  The $\gamma$-ray
spectrum for  this process for power-law and monoenergetic electrons 
scattering thermal blackbody X-ray photons above the neutron star
surface (Daugherty \& Harding 1989) is a power-law with maximum energy 
$\erg_{max} \simeq \gamma_c B' = 2 \times 10^3\,B'^2/T_6$ (Dermer 1990),
where $\gamma_c$ is the energy above which the electrons scatter 
resonantly and $T_6 \equiv T/10^6$ K is the thermal X-ray temperature.
In the case of PSR1509-58 with $B' = 0.7$, $\erg_{max} \simeq 10^3/T_6$. 
Since the thermal surface emission component is not observed due to the
strong non-thermal spectrum seen at X-ray energies (Kawai 1993), $T_6$
is not known.  However, PSR1509-58 is young ($\sim 1000$ yr) and
probably has $T_6 \sim 1 - 3$.   We would then expect $\erg_{max} \simeq
300 - 10^3$, compatible with our choice of $\erg_{max} = 100$ for the
splitting cascade models.

A dozen or so other radio pulsars have spin-down magnetic fields above 
$10^{13}$ G.  These pulsars would, like PSR1509-58, have photon splitting
dominated cascades rather than pair cascades, producing lower yields of 
electron-positron pairs.  It is possible that neutron stars with extremely
high magnetic fields, where splitting is dominant at altitudes up to
several stellar radii, do not produce sufficient pairs for coherent radio
emission, an intriguing possibility.  If such neutron stars exist, they
would constitute a new class of radio quiet, low-energy $\gamma$-ray pulsars.

\acknowledgements
We thank Dieter Hartmann and David Thompson for reading the manuscript
and for providing helpful comments, and Wim Hermsen, Alberto Carrami\~nana 
and Kevin Bennett for providing preliminary Comptel data for PSR1509-58.
This work was supported through Compton Gamma-Ray Observatory Guest 
Investigator Phase 5 and NASA Astrophysics Theory Grants.
MGB thanks the Institute for Theoretical Physics at the University of
California, Santa Barbara for support (under NSF grant PHY94-07194)
during part of the period in which work for this paper was completed.

\clearpage

\appendix

\section{CURVED SPACETIME EFFECTS}

We include here some details of our treatment of general relativistic effects
on the photon splitting and pair production attenuation in a  neutron star
magnetosphere.  This treatment follows closely that of Gonthier \& Harding
(1994), who examined the importance of general relativistic effects on
one-photon pair production attenuation in a  Schwarzschild metric.  They found
that several effects of curved spacetime make significant corrections to the
attenuation lengths and escape energies for this process, namely the curvature
of the photon trajectories, the redshift of the photon energy, and the change
in the dipole magnetic field.  In the present calculation, the first two
effects are included together in the expression for the photon momentum
4-vector.  The curved trajectory of a photon in the Schwarzschild metric is
confined to a single plane, which we may specify as the x-z plane.  For an
observer at rest in the local inertial frame at a radius \teq{r} in the
Schwarzschild field, the components of the momentum 4-vector are then
\begin{eqnarray}  
   k^{\hat r} & = & \left[(1 - 2u)^{-1}-{u^2\over u_b^2}\right]^{1/2}\,\erg 
   \nonumber \\
   k^{\hat{\vartheta}} & = & {u\over u_b}\,\erg \nonumber \\
   k^{\hat{\phi}} & = & 0 \nonumber \\
   k^{\hat o} & = & (1 - 2u)^{-1/2}\erg\quad ,      \label{eq:4momgr}
\end{eqnarray}
where \teq{u = m/r}, with \teq{m = GM/c^2} as the scaled stellar mass
(i.e. the Schwarzschild radius), 
\teq{\erg} is the photon energy as observed at infinity, 
\begin{equation}
   u_b = \dover{m}{R\sin\delta_o}\,\sqrt{1-\dover{2m}{R}}
\end{equation}
for a neutron star radius of \teq{R}, and $\delta_o$ is the initial
propagation angle of the photon to the radial direction.  The
\teq{\hat\vartheta} and \teq{\hat o} components of \teq{k} are adapted
from equations~(13) and~(14) of Gonthier \& Harding (1994),
\teq{k^{\hat\phi}=0} follows from the choice of the plane of
propagation, and the \teq{\hat r} component can be deduced from the
others using  \teq{k_\mu k^\mu =0}. 

At the photon emission point, we first determine the angle $\delta_o$
by performing two coordinate transformations to put the photon momentum
3-vector in the x-z plane.  The spacetime trajectory for that photon is
then computed and stored in two tables: the first giving the value of $r$
as a function of the total pathlength along the trajectory, $s = c\Delta\tau$,
from the time-of-flight in the local frame,
\begin{equation}
   \Delta \tau\; =\; -\dover{m}{c}\int^{m/r}_{m/R}{u_b\, du \over
   {\left[u_b^2-u^2(1-2u) \right]^{1/2} (1-2u)^{1/2} u^2}} \quad , 
  \label{eq:itime}
\end{equation}
which closely resembles Eq.~(19) of Gonthier and Harding (1994, which
instead measures time in the non-local observer's frame), and the second 
table giving the value of $\vartheta$ as a function of $r$ from the
equation,
\begin{equation}
   \left({du\over {d\vartheta}}\right)^2=u_b^2-u^2(1-2u)\quad . 
  \label{eq:traject}
\end{equation}
The $r$ and $\vartheta$ are the coordinates in the x-z plane of each
point  along the photon trajectory. At each distance $s$ along the
photon path from the emission point, we look  up the value of $r$, and
from $r$ deduce the value of $\vartheta$.  These values of $r$ and
$\vartheta$ then define the new position and momentum 4-vectors in the
x-z plane.  The inverse coordinate transformations of those described
above will then give the position and momentum 4-vectors in the frame in
which we carry out the attenuation length calculation.

To describe the magnetic field in curved spacetime, we use the expression
of Wasserman \& Shapiro (1983) for the dipole field measured in the local
inertial frame in a Schwarzschild metric:
\begin{eqnarray}
   \vec{B}_{curved} & = & - {3\over 2}{B_0\cos\vartheta\over {m^2 r}}
   \left[{r\over{2m}}\ln\left(1-{2m\over r}\right)
   + 1 + {m\over r}\right]\hat{r} + \nonumber \\
   & & {3\over 2}{B_0\sin\vartheta\over {m^2 r}}\left[\left({r\over
   {2m}}-1\right)\ln\left(1-{2m\over r}\right)+1-{m\over r}\right]
   \left(1-{2m\over r}\right)^{-1/2}\hat{\vartheta}\quad . \label{eq:bcurv}
\end{eqnarray}
Note that this expression in GH94 has a typographical error in the $\hat
{\vartheta}$ component.  For $M = 1.4 M_{\odot}$ and $R = 10$ km, as used
in this paper, the dipole field strength at the neutron star surface at
the pole from equation~(\ref{eq:bcurv}) is a factor $\sim 1.4$ times the flat 
space value.

The angle $\theta_{\rm kB}$, obtained by taking the dot product between the 
photon momentum 4-vector and the local dipole field, is given in
curved spacetime by
\begin{equation}
   \cos\theta_{\rm kB} = {B^{\hat{r}} \over B} \left[1-(1-2u){u^2\over u^2_b}
   \right]^{1/2} + {B^{\hat{\vartheta}} \over B} {(1-2u)^{1/2}u\over u_b}\; .
  \label{eq:thetakb}
\end{equation}
Since the field components \teq{B^{\hat{r}}} and \teq{B^{\hat{\vartheta}}}
defined by \teq{\vec{B}_{curved}=B^{\hat{r}} \hat{r} + B^{\hat{\vartheta}}
\hat{\vartheta}} in equation~(\ref{eq:bcurv}) are differentiable
functions of \teq{u=m/r}, it follows from equation~(\ref{eq:thetakb})
that \teq{\theta_{\rm kB}} is also differentiable in \teq{r}.  Note also
that the pathlength \teq{s} defined through equation~(\ref{eq:itime})
gives smooth \teq{\partial s/\partial r} at all positions above the
stellar surface.  Consequently, the implicit function \teq{\theta_{\rm kB}(s)} 
has a well-behaved Taylor series expansion (i.e. there is no singularity)
about \teq{r=R} (i.e. \teq{s=0}).  Hence, cases where we set
\teq{\theta_{\rm kB,0}=0} result in \teq{\theta_{\rm kB}\propto s} along
the photon trajectory for \teq{(r-R)/R\ll 1}, regardless of whether the
spacetime is curved or flat.  This proportionality is responsible for
certain limiting behaviors in the attenuation lengths and escape
energies discussed in Sections~2.3 and~2.4.

\clearpage
~
\vskip -2.5in
\figureout{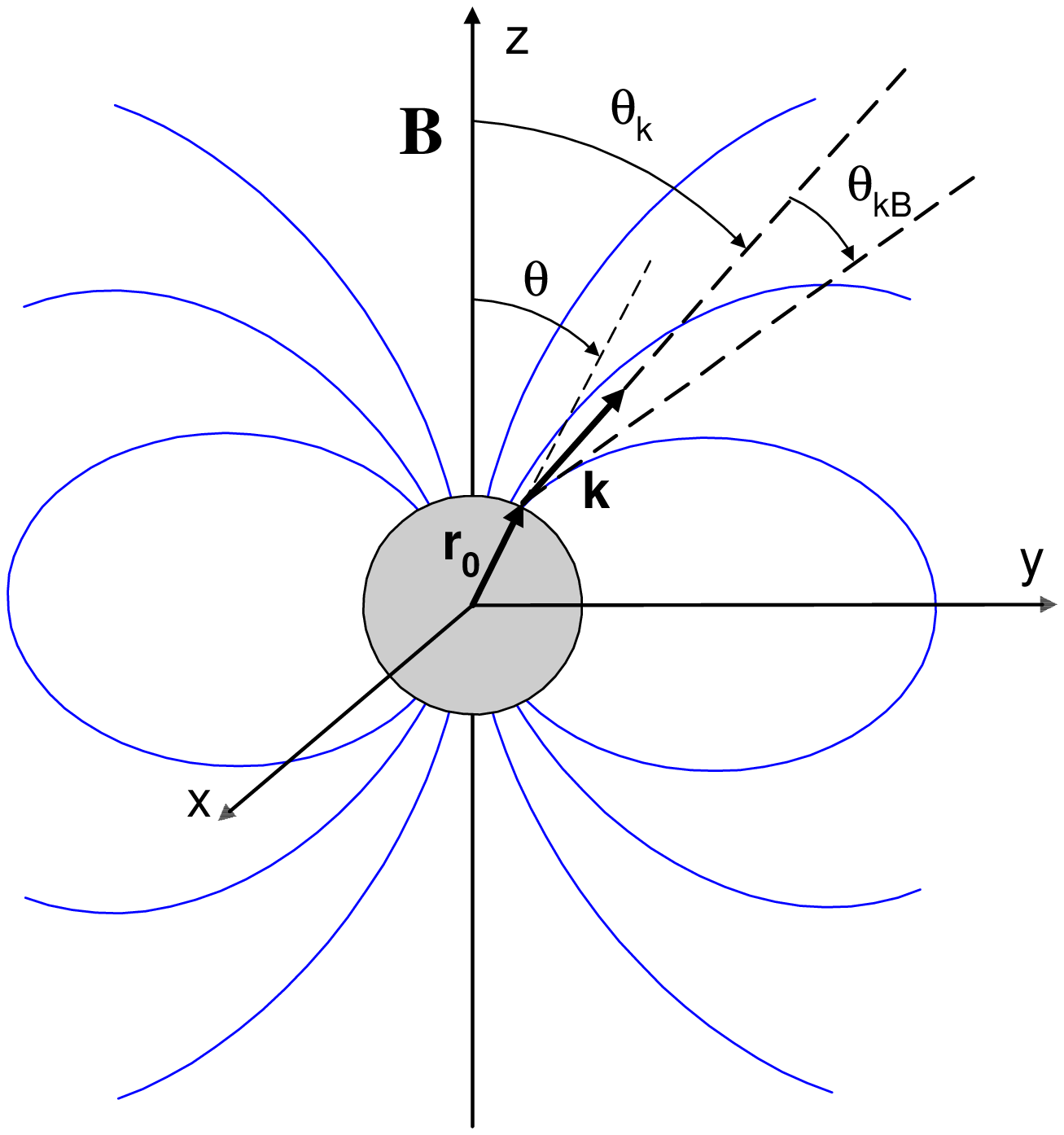}{0}{
 Schematic illustration of the neutron star dipole magnetic field geometry, used
 for determination of attenuation lengths and escape energies.  The dipole field
 has an axis in the z-direction, and the photon originates at position vector
 \teq{{\bf r}_{\bf 0}} on the neutron star surface, labelled by the polar angle
 \teq{\theta}.  The photon propagates in the direction of its momentum vector
 \teq{{\bf k}} that makes an angle \teq{\theta_{\rm kB}} to the
 local field ($\theta_{\rm kB,0}$ at the emission point), and is 
 described by polar angle \teq{\theta_k}  with respect to
 its original location \teq{{\bf r}_{\bf 0}}.  For all  results in this paper,
 we arbitrarily choose a photon trajectory in  x-z plane, corresponding to a
 phase \teq{\phi = 0} (see the Appendix).
   \label{fig:geometry}}         

\figureout{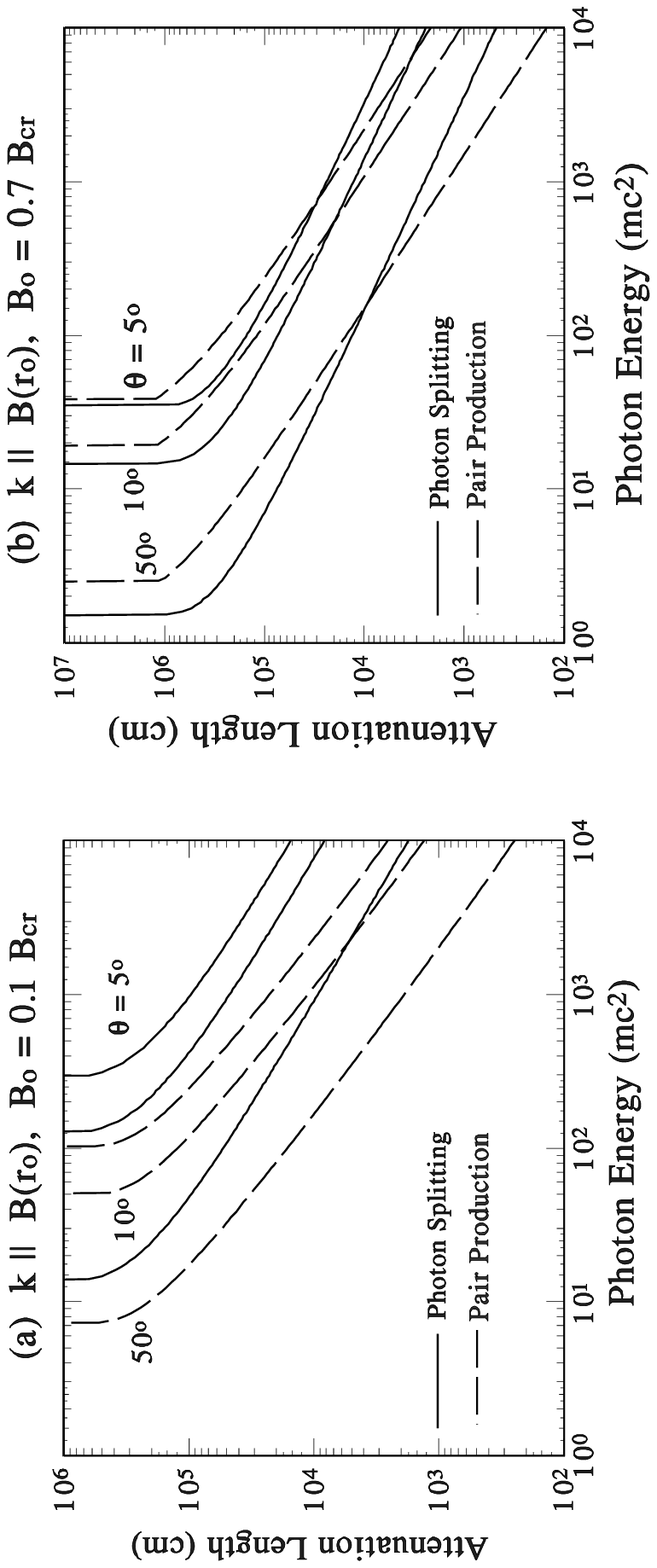}{270}{
 The attenuation length \teq{L} for photon splitting, defined in
 equation~(\protect\ref{eq:splitotrate}) assuming three CP-permitted modes, 
 and for single photon pair production as a function of energy for photons 
 initially propagating parallel to the local field (i.e. 
 \teq{\theta_{\rm kB,0}=0}), at different colatitudes $\theta$ on the
 neutron star surface.  Two cases are depicted, namely for surface fields 
 of (a)
 \teq{B_0=0.1B_{\rm cr}} and (b) \teq{B_0=0.7B_{\rm cr}}, the latter being the
 spin-down field strength for PSR1509-58.  At high energies \teq{\erg}, the
 lengths scale as \teq{\erg^{-5/7}} for 
 photon splitting and \teq{\erg^{-1}} for
 pair production, as discussed in the text.  The lengths are averaged over
 photon polarizations.
   \label{fig:attenl}}       

\figureout{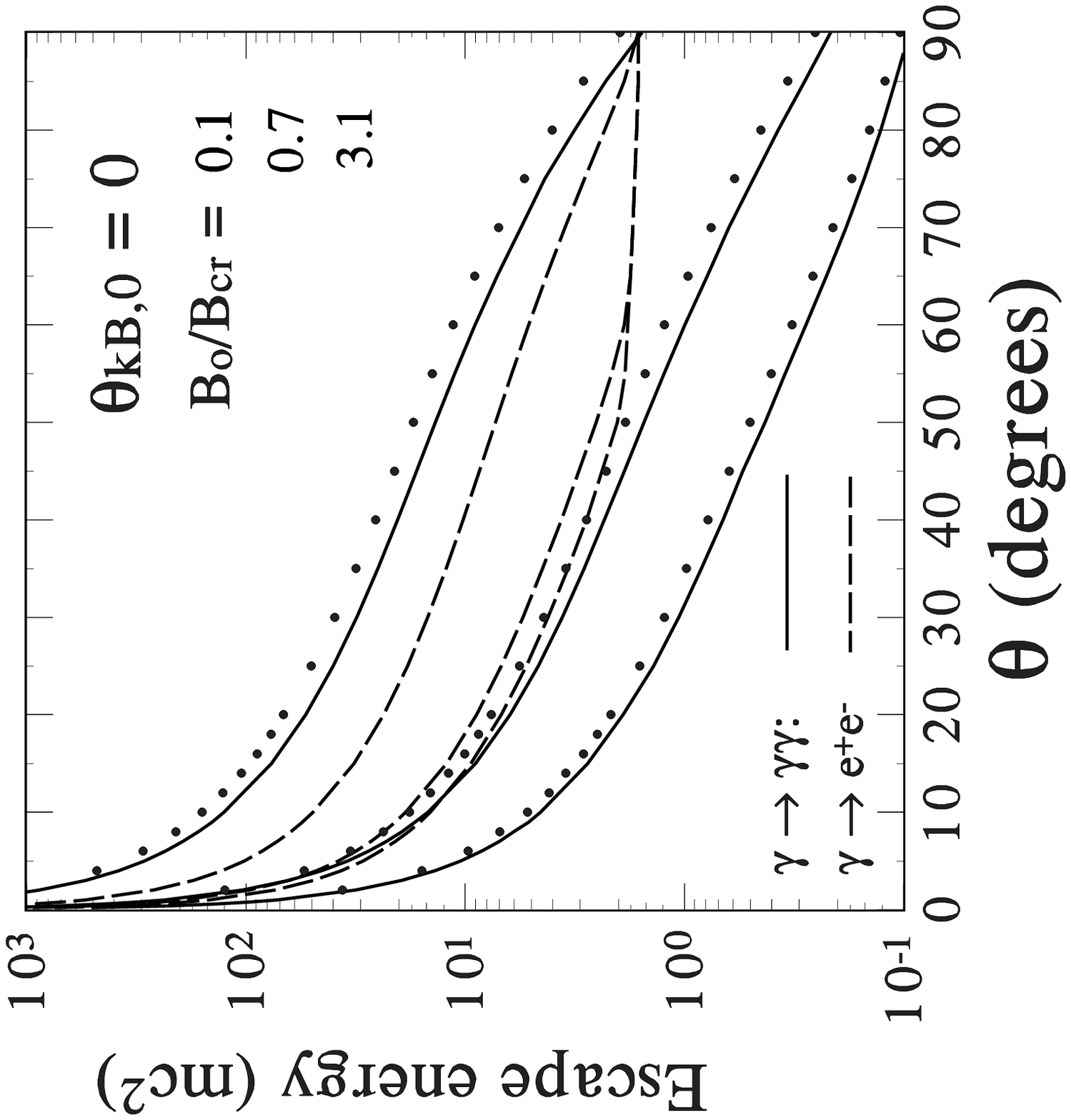}{270}{
 The escape energy (i.e. where \teq{L\to\infty} in Eq.~[\ref{eq:tau}]),
 in units of $m_ec^2$, for photon splitting, averaged over all modes (solid     
 curves) and in the $\perp \rightarrow \parallel\parallel$ mode only (solid
 dots), 
 compared to the escape energy for one-photon pair production (dashed curves),
 both as functions of magnetic colatitude $\theta$ of the emission point on the
 neutron star surface.  These energies are obtained for three different 
 surface dipole magnetic field
 strengths and for emission along the field ($\theta_{\rm kB,0}=0$).  The curves
 diverge near $\theta = 0$, where the field line radius of curvature becomes
 infinite; these divergences scale as \teq{\theta^{-6/5}} for splitting and
 \teq{\theta^{-1}} for pair production (see text and also 
 Fig.~\ref{fig:icescape} below).  The escape energies for each process are
 monotonically decreasing functions of \teq{B} for the range of parameters
 shown.  The escape energies are averaged over photon polarizations and
 computed using the Schwarzschild metric.
   \label{fig:escape} }    

\figureout{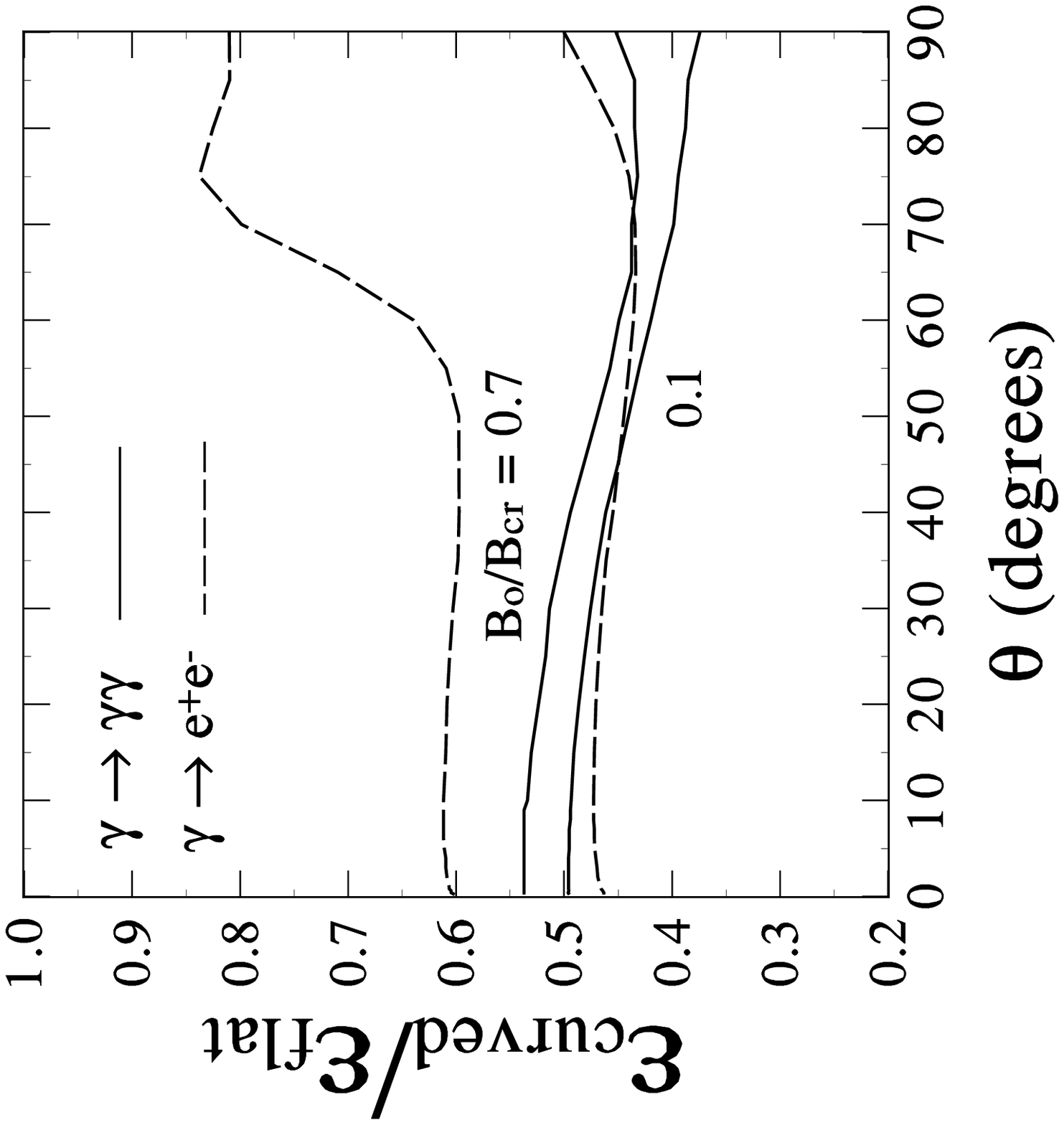}{270}{
 The ratio of escape energies in curved and flat spacetime for photon splitting
 and pair production, for emission parallel to the field as a function of 
 magnetic colatitude \teq{\theta} of the
 emission point.
    \label{fig:escurvflat} }    

\figureout{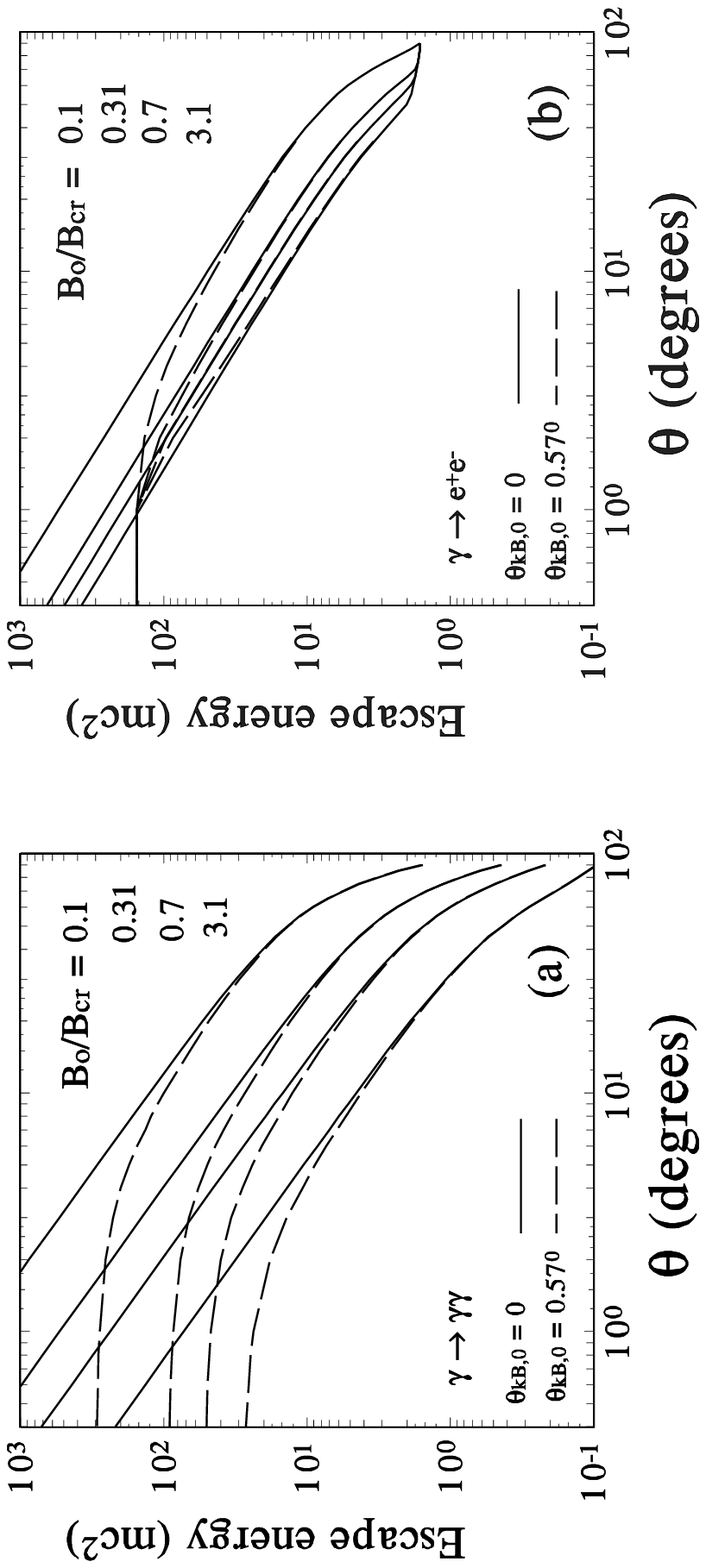}{270}{
 The escape energy for (a) photon splitting and (b) pair production  as a
 function of magnetic colatitude for photon emission both along ${\bf B}$
 and at angle $\theta_{\rm kB,0}=0.01$ radians (\teq{=0.57^\circ}) to the
 field.  At low magnetic colatitudes \teq{\theta\lesssim 10\theta_{\rm
 kB,0}}, the field curvature is so low that photon attenuation is
 insensitive to the value of \teq{\theta} and  is well described by the
 uniform field results in  equations~(\ref{eq:splitotrate})
 and~(\ref{eq:ppratlim}--\ref{eq:xi}). The \teq{\theta_{\rm kB,0}=0}
 (solid) curves have slopes of -6/5 (splitting) and -1 (pair creation) at
 small \teq{\theta}, as discussed in the text.
    \label{fig:icescape} }    

\figureout{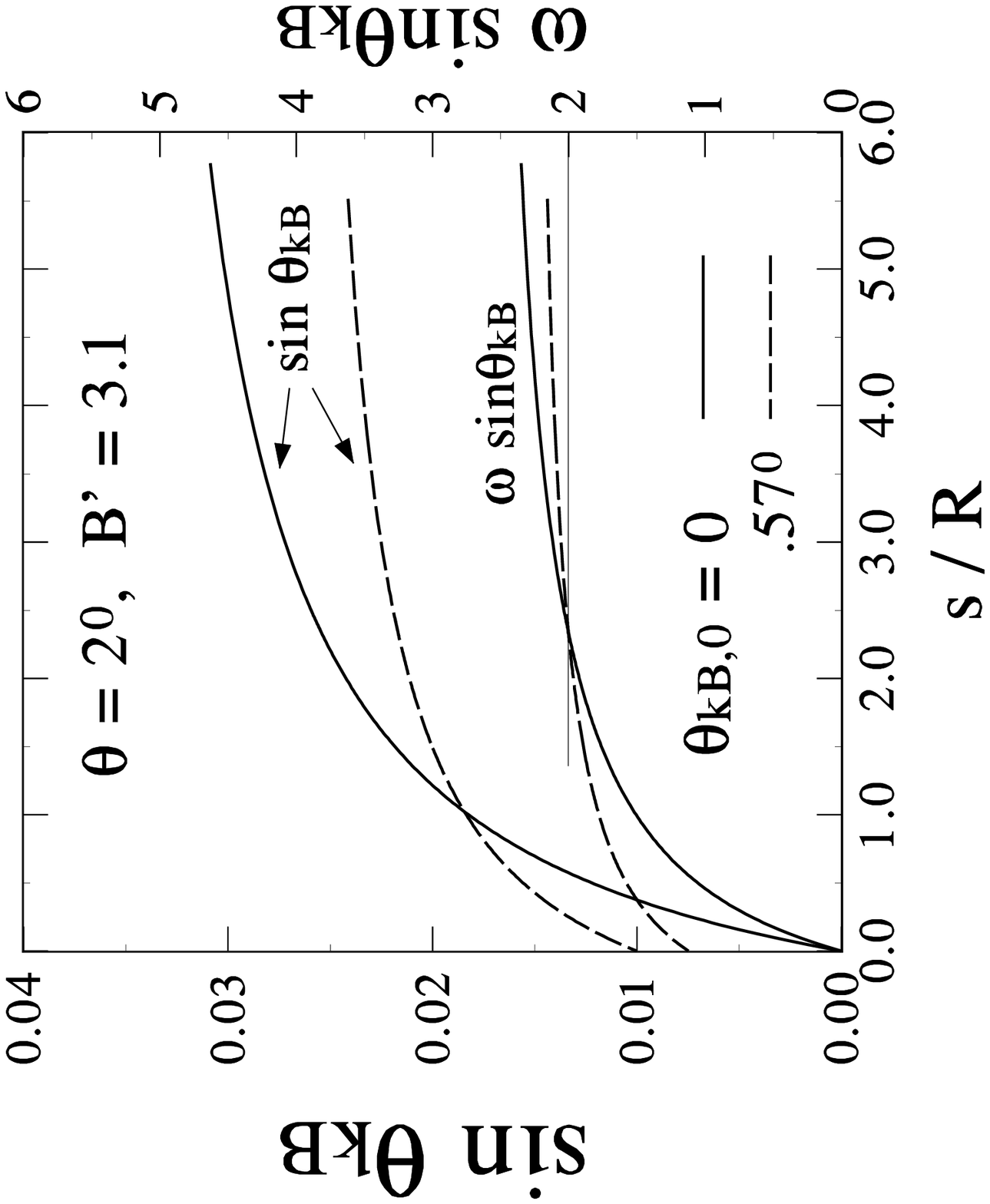}{270}{
 The variation of $\sin\theta_{\rm kB}$ (left-hand scale) and 
 $\omega\sin\theta_{\rm kB}$ (right-hand scale) with
 path length $s$ above the neutron star surface, scaled by the radius $R$, in
 curved spacetime for two different values of $\theta_{\rm kB,0}$.  
 $\omega$ is the photon energy in the local inertial frame and the
 light solid horizontal line marks the pair threshold.  Observe that
 \teq{\theta_{\rm kB}\propto s} for \teq{s/R\ll 1}.  The colatitude
 \teq{\theta=2^\circ} and field strength \teq{B'=3.1} are chosen 
 specifically to facilitate the understanding of Fig.~\ref{fig:icescape}.
    \label{fig:wsinthet} }    

\figureout{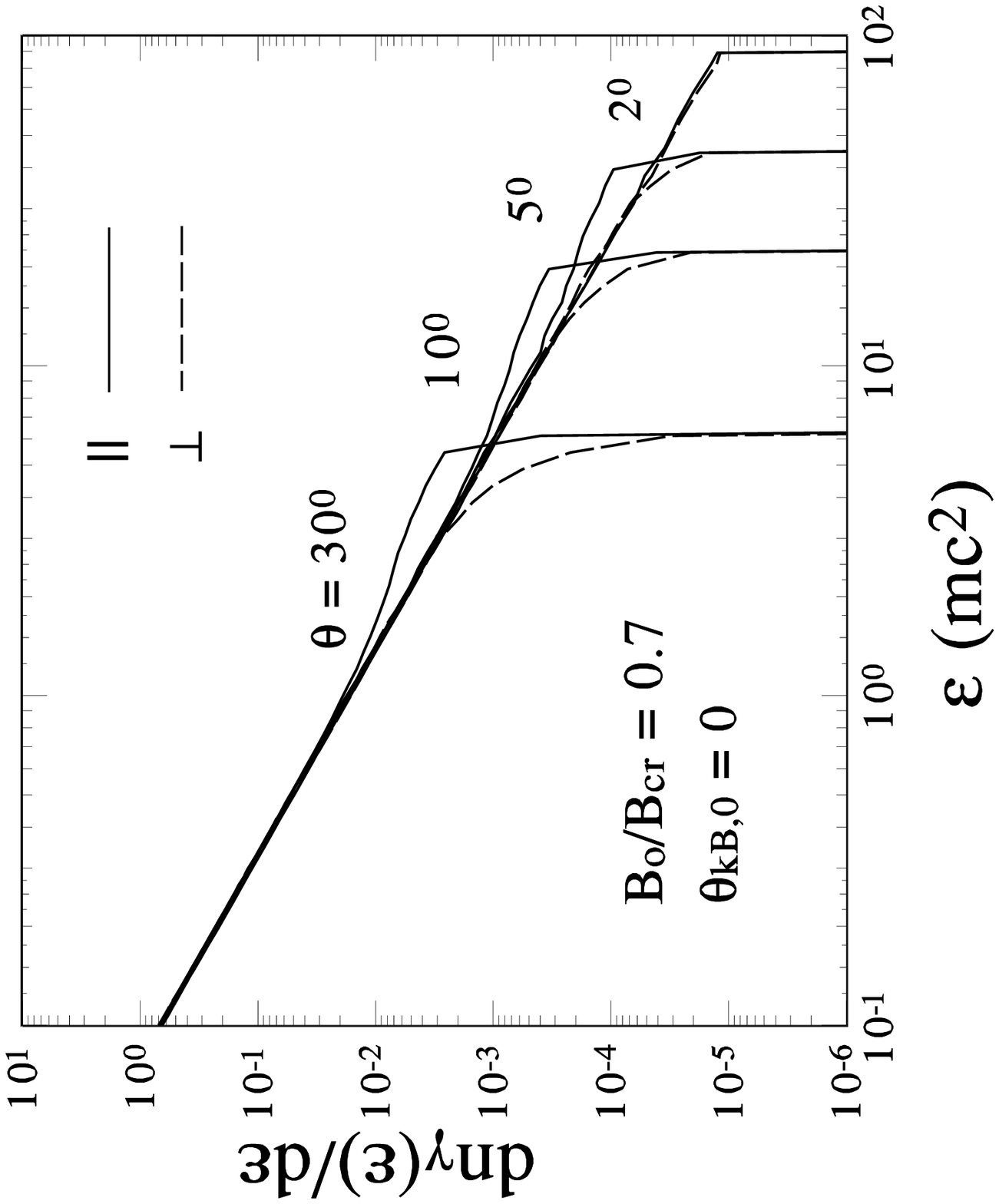}{270}{
 Polarized spectra for partial photon splitting cascades, assuming
 unpolarized power-law emission (of index \teq{\alpha =1.6}) parallel to
 the magnetic field ($\theta_{\rm kB,0} = 0$), at different magnetic
 colatitudes, $\theta$, as labelled.  Here only photons of polarization
 \teq{\perp} split, while those of either polarization produce pairs.
 The normalization of the spectrum is arbitrary.   
    \label{fig:specpar} }    

\figureout{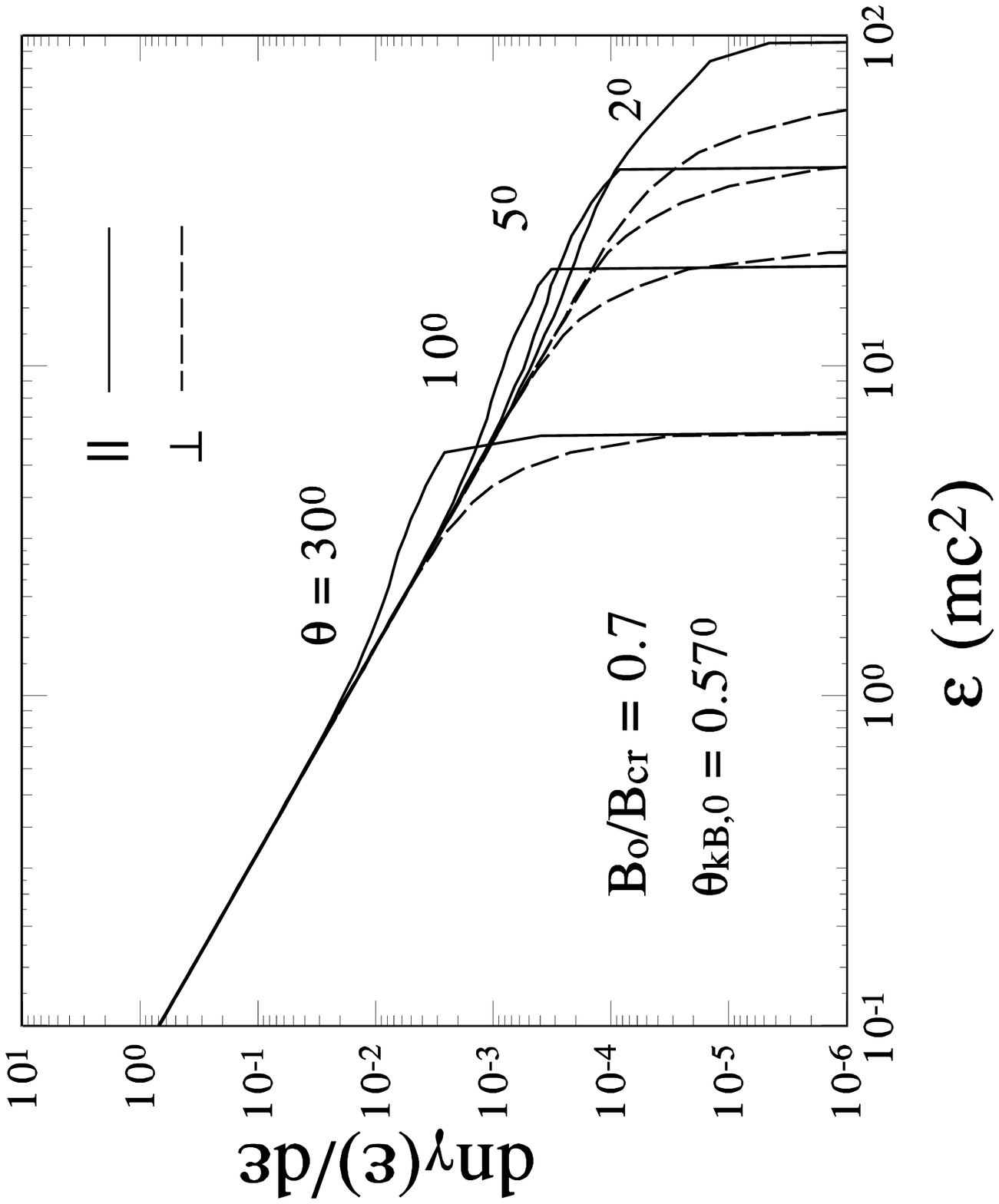}{270}{
 Same as Fig.~\ref{fig:specpar}, but for emission at angle 
 $\theta_{\rm kB,0} = 0.01$ radians (\teq{=0.57^\circ}) to the field. 
 Spectra differ only marginally from Fig.~\ref{fig:specpar}.
    \label{fig:icspecpar} }    

\figureout{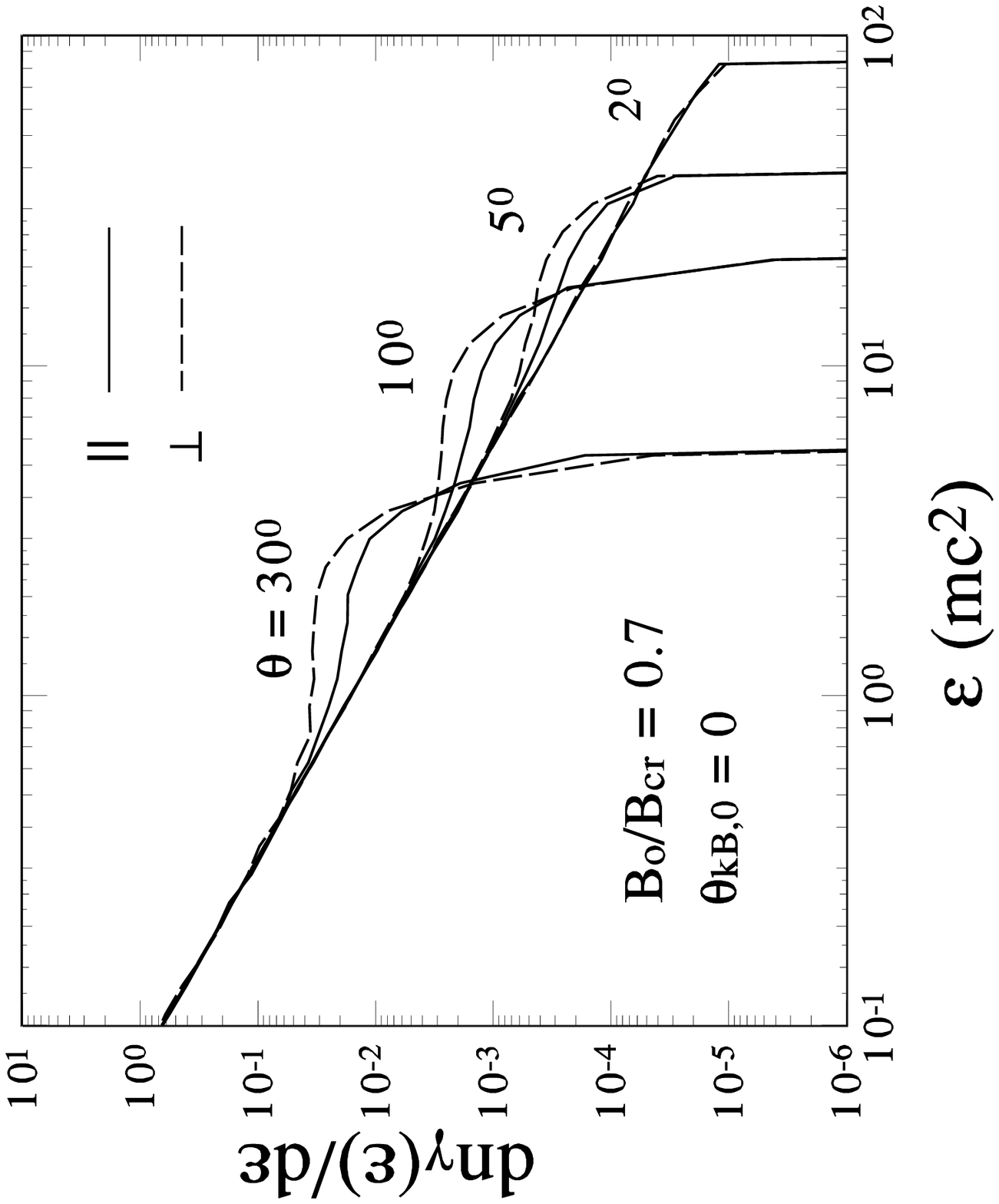}{270}{
 Polarized spectra for full photon splitting cascades, assuming
 unpolarized power-law emission (of index \teq{\alpha =1.6}) parallel
 to the magnetic field ($\theta_{\rm kB,0} = 0$), at different magnetic
 colatitudes, $\theta$.  The cutoffs occur at energies comparable to the
 escape energies computed in Section 2.4.  Pair creation is permitted
 in these runs, and is generally small away from the pole.  
   \label{fig:specpol} }    

\figureout{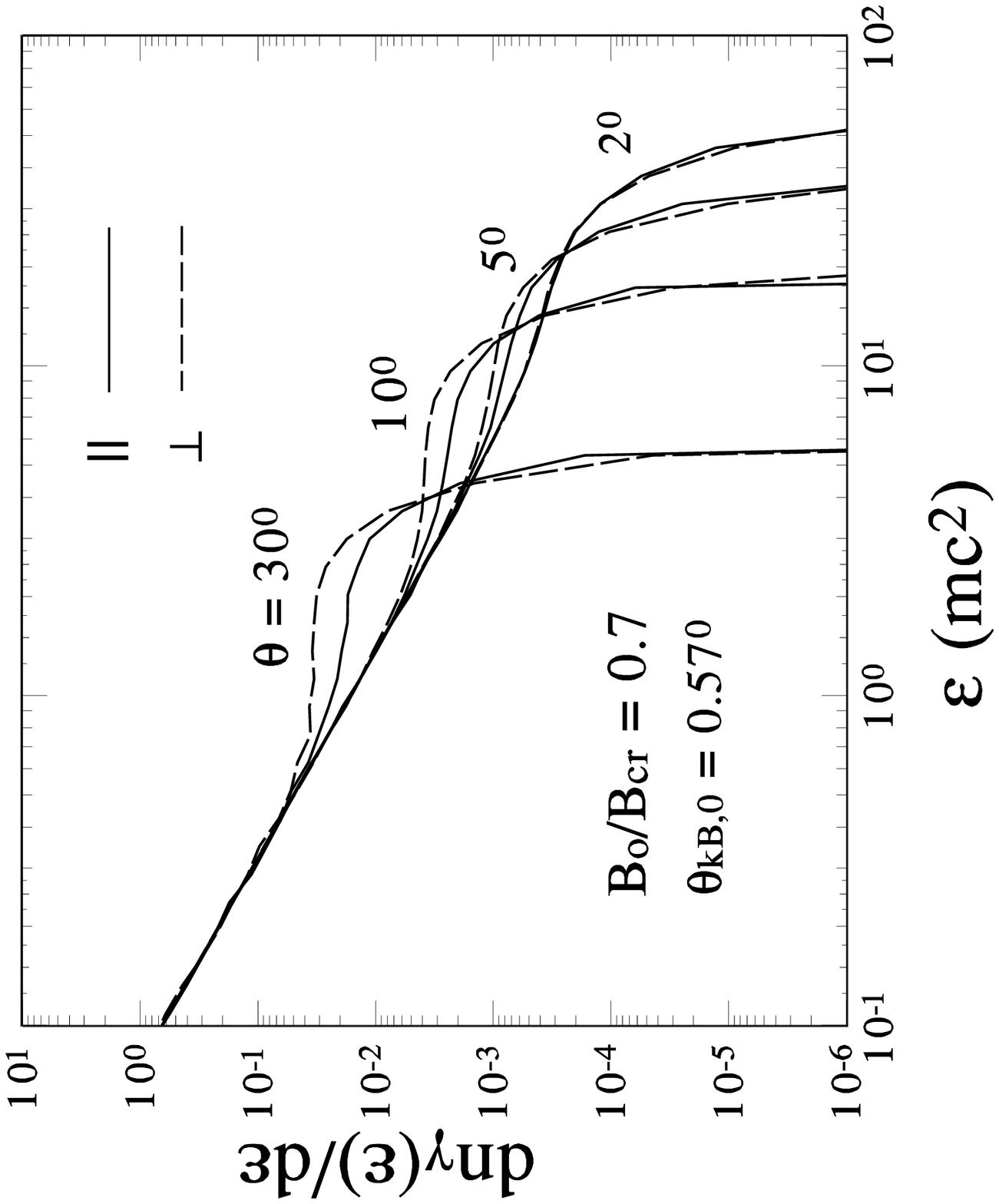}{270}{
 Same as Fig.~\ref{fig:specpol}, but for emission at angle $\theta_{\rm
 kB,0} = 0.57^\circ$ to the field, towards the dipole axis.  This
 explores the effect  of finite opening angle of emission, namely that
 the attenuation is considerably more severe than in
 Fig.~\ref{fig:specpol} at colatitudes close to the pole.
   \label{fig:icspecpol} }    

\figureout{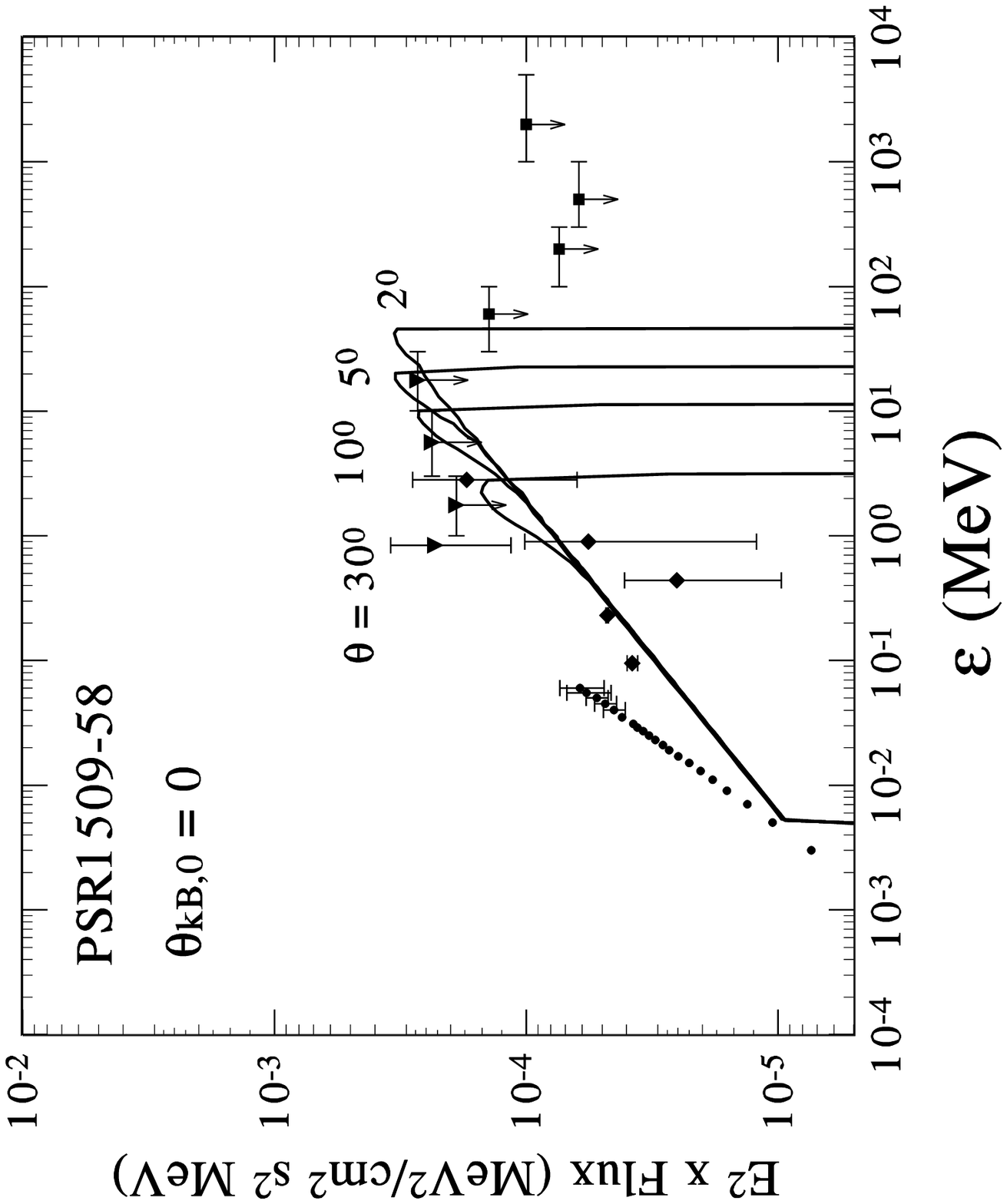}{270}{
 Partial photon splitting cascade spectra, obtained by averaging the
 spectra from Fig.~\ref{fig:specpar} over polarizations and multiplying
 them by \teq{\erg^2}, compared to the observed spectrum from PSR1509-58.
 Data points are from GINGA (Kawai, Okayasu, and Sekimoto, 1993): circles,
 OSSE (Matz et al. 1994): diamonds, COMPTEL  
 (Hermsen et al. 1996): triangles, and EGRET (Nel et al. 1996): 
 squares, and the collective display is an updated version of that 
 in Thompson (1996).
    \label{fig:datpar} }    

\figureout{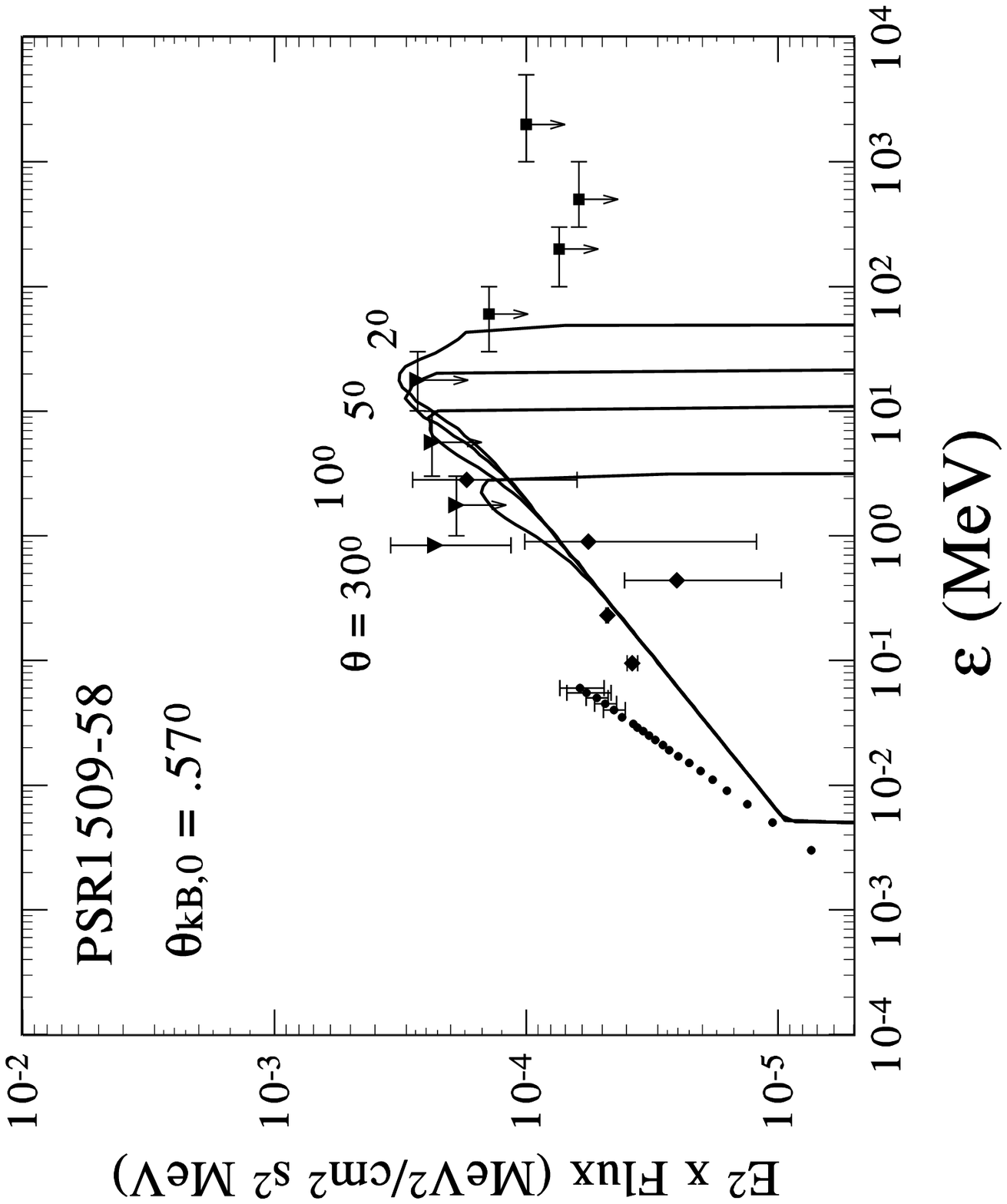}{270}{
 Same as Fig.~\ref{fig:datpar}, for the model spectra of
 Fig.~\ref{fig:icspecpar}.
    \label{fig:daticpar} }    

\figureout{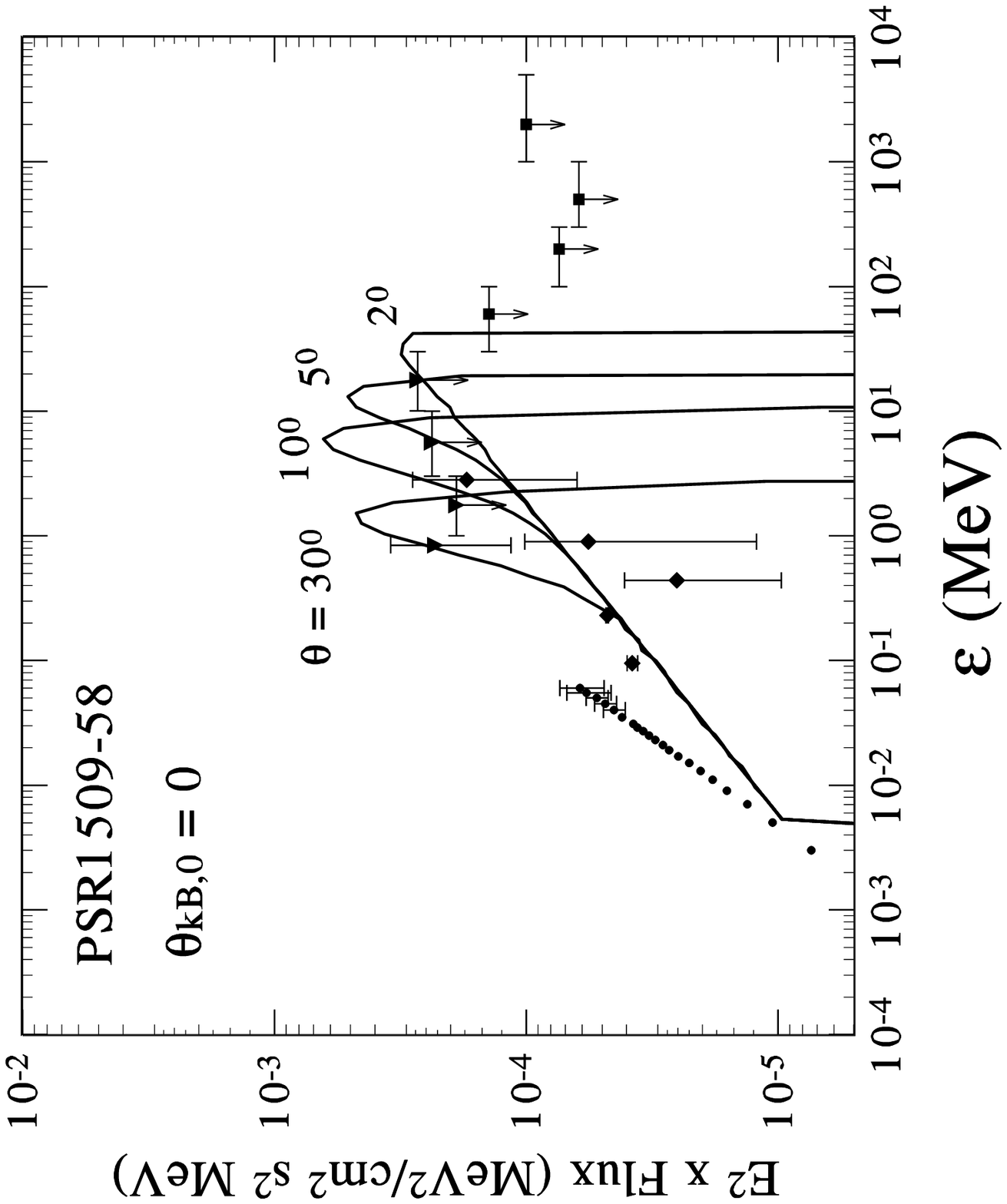}{270}{
 The equivalent of Fig.~\ref{fig:datpar} for full splitting cascades, 
 i.e. obtained by averaging the model spectra of Fig.~\ref{fig:specpol}
 over polarizations and multiplying them by \teq{\erg^2}.
    \label{fig:datunpol} }    

\figureout{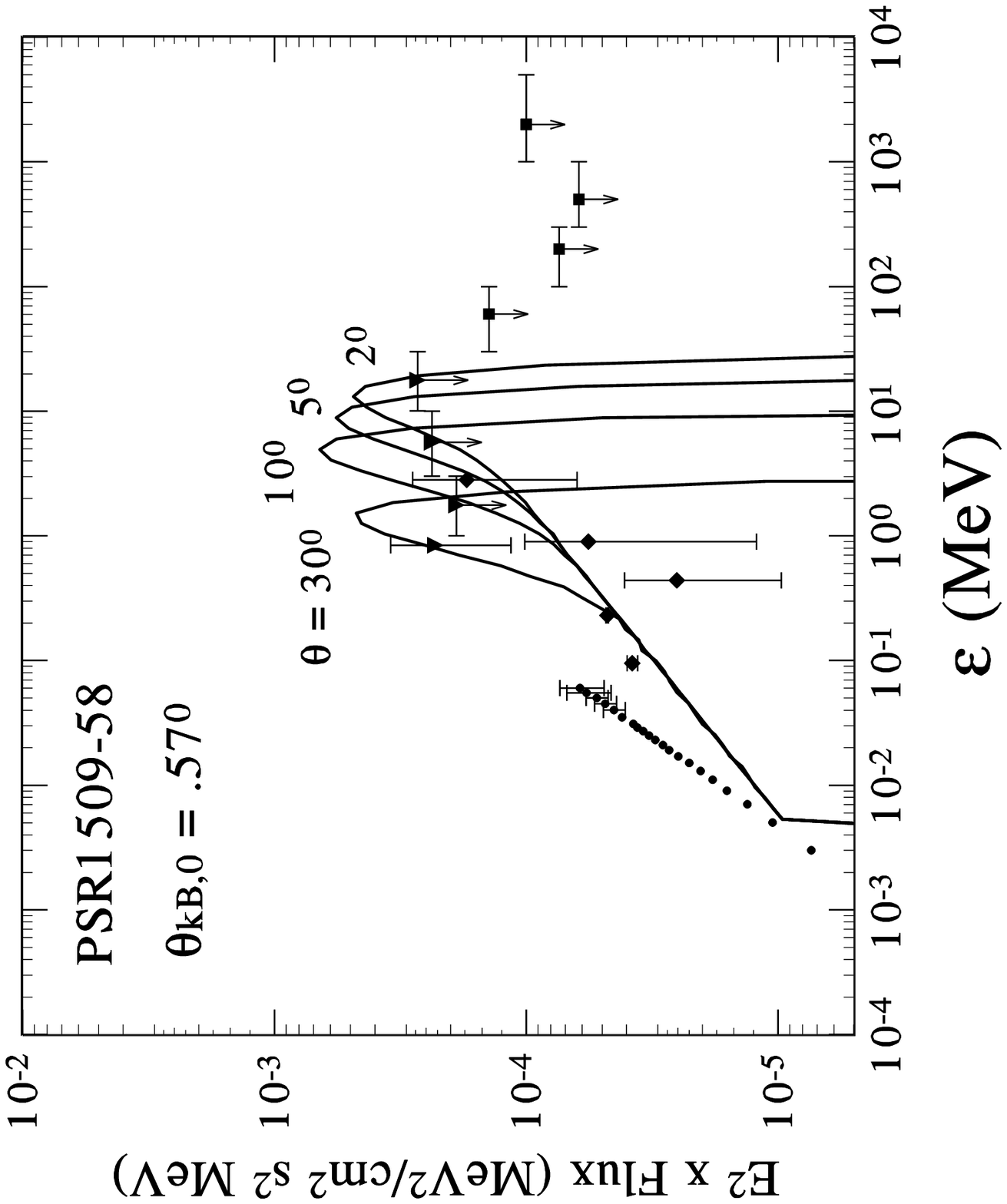}{270}{
 Same as Fig.~\ref{fig:datunpol}, for the model spectra of
 Fig.~\ref{fig:icspecpol}.
    \label{fig:daticunpol} }   


\clearpage

\begin{thebibliography}{}

\bibitem{}
   Adler, S.~L., 1971, Ann. Phys.,67, 599.
\bibitem{}
   Adler, S.~L., Bahcall, J.~N., Callan, C.~G., \& Rosenbluth, M.~N.: 1970
   \prl, 25, 1061. 
\bibitem{}
   Arons, J. 1983, \apj, 266, 215.
\bibitem{}
   Baring, M. G. 1988, \mnras, 235, 79. 
\bibitem{}
   Baring, M.~G.: 1991 \aap, 249, 581.
\bibitem{}
   Baring, M.~G.: 1993 \mnras, 262, 20.
\bibitem{}
   Baring, M. G. 1995, \apjl, 440, L69.
\bibitem{}
   Baring, M. G. \& Harding, A. K., 1995a, \apss, 231, 77.
\bibitem{}
   Baring, M. G. \& Harding, A. K., 1995b, Proc. of 24th Intl. 
   Cosmic Ray Conf., Vol. II, p. 271.
\bibitem{}
   Baring, M. G. \& Harding, A. K., 1996, ApJ Letters, submitted.
\bibitem{}
   Ba\u{\i}er, V.~N., Mil'shte\u{\i}n, A.~I., \&
   Sha\u{\i}sultanov, R.~Zh 1986, Sov. Phys. JETP, 63, 665.
\bibitem{}
   Ba\u{\i}er, V.~N., Mil'shte\u{\i}n, A.~I., \&
   Sha\u{\i}sultanov, R.~Zh 1996. \prl, 77, 1691.
\bibitem{}
   Bennett, K. et al., 1994, \apjs, 90, 823.
\bibitem{}
   Bertsch, D. L. et al. 1992, \nat, 357, 306.
\bibitem{}
   Bialynicka-Birula, Z., \& Bialynicki-Birula, I. 1970, \prd, 2, 2341.
\bibitem{}
   Bulik, T. \& Miller, M. C. 1996, preprint.
\bibitem{}
   Chang, H.-K., Chen, K., Fenimore, E. E., \& Ho., C. 1996, Proc. Huntsville
   Workshop on Gamma-Ray Bursts, (AIP, New York) in press. 
\bibitem{}
   Chang, H.-K., Chen, K., \& Ho., C. 1996, \aap, in press. 
\bibitem{}
   Cheng, K. S., Ho, C. \& Ruderman, M. A. 1986, \apj, 300, 500.
\bibitem{}
   Chiang, J. \& Romani, R. W. 1992, \apj, 400, 629.
\bibitem{}
   Daugherty, J.~K. \& Harding A.~K., 1982, \apj, 252, 337.
\bibitem{}
   Daugherty, J. K. \& Harding, A. K. 1983, \apj, 273, 761. (DH83)
\bibitem{}
   Daugherty, J.~K. \& Harding A.~K., 1994, \apj, 429, 325.
\bibitem{}
   Daugherty, J.~K. \& Harding A.~K., 1996, \apj, 458, 278.
\bibitem{}
   Fierro, J. M. et al. 1993, \apjl, 413, L27.
\bibitem{}
   Gonthier, P. L.  \& Harding, A. K., 1994, \apj, 425, 767.
\bibitem{}
   Halpern, J.  P. \& Holt, S. S. 1992, \nat, 357, 222.
\bibitem{}
   Harding, A. K. 1995, in Millesecond Pulsars: A Decade of Surprises,
   ed. A. S. Fruchter, M. Tavani \& D. C. Backer (ASP Conf. Proc. Vol. 72), 
   p. 322. 
\bibitem{}
   Harding, A. K. \& Baring, M. G. 1996, Proc. Huntsville
   Workshop on Gamma-Ray Bursts, (AIP, New York) in press. 
\bibitem{}
   Harding, A. K., Baring, M. G. \& Gonthier, P. L. 1996, \aap, in press.
\bibitem{}
   Hartmann, D. H. et al., 1993, in {\it Isolated Pulsars}, 
   ed. K. van Riper, R.~Epstein \& C. Ho, Cambridge Univ. Press, p. 405.
\bibitem{}
   Hermsen, W. et al. 1996, Adv. Space Research, in press.
\bibitem{}
   Jauch, M.~M., \& Rohrlich, F. 1980, The Theory of Photons 
   and Electrons (2nd edn. Springer, Berlin).   
\bibitem{}
   Kanbach, G. et al. 1994, \aap, 289, 855.
\bibitem{}
   Kawai, N., Okayasu, R. \& Sekimoto, Y. 1993, in Compton Gamma-Ray
   Observatory, eds. M. Friedlander, N. Gehrels, D.~J. Macomb 
   (AIP Conf. Proc. 280, AIP, New York) p.~213 
\bibitem{}
   Klepikov, N. V. 1954, Zh. Eksp. Theor. Fiz., 26, 19.
\bibitem{}
   Manchester, R. N. \& Taylor, J. H., 1977, Pulsars (Freeman, 
   San Francisco).
\bibitem{}
   Matz, S. M. et al. 1994, \apj, 434, 288.
\bibitem{}
   Mayer-Hasselwander, et al. 1994, \apj, 421, 276.
\bibitem{}
   Melrose, D.~B., \& Parle, A. J. 1983a, Aust. J. Phys., 36, 775
\bibitem{}
   Melrose, D.~B., \& Parle, A. J. 1983b, Aust. J. Phys., 36, 799
\bibitem{}
   Mentzel, M., Berg, D \& Wunner, G. 1994, Phys. Rev. D, 50, 1125.
\bibitem{}
   Michel, F. C., 1991, Theory of Neutron Star Magnetospheres
   (Univ. of Chicago Press).
\bibitem{}
   Minguzzi, A. 1961, Nuovo Cimento, 19, 847.
\bibitem{}
   Mitrofanov, I.~G., Pozanenko, A.~S., Dolidze, V.~Sh., Barat, C., 
   Hurley, K., Niel, M., \& Vedrenne, G. 1986, Sov. Astron, 30, 659. 
\bibitem{}
   Nel, H. I. et al. 1992, \apj, 398, 602.
\bibitem{}
   Nel, H. I. et al. 1996, \apj, in press.
\bibitem{}
   Nolan, P. L. et al. 1993, \apj, 409, 697.
\bibitem{}
   Papanyan, V. O. \& Ritus, V. I., 1972, Soviet JETP, 34, 1195.
\bibitem{}
   Ramanamurthy, P. V. et al. 1995, \apjl, 447, L109.
\bibitem{}
   Ramanamurthy, P. V. et al. 1996, \apj, in press.
\bibitem{}
   Romani, R. W. \& Yadigaroglu, I.-A. 1995, \apj, 438, 314.
\bibitem{}
   Ruderman, M. A. \& Sutherland, P. G. 1975, \apj, 196, 51.
\bibitem{}
   Shabad, A. E. 1975, Ann. Phys., 90, 166.
\bibitem{}
   Shapiro, S. L. \& Teukolsky, S. A., 1983, Black Holes, White
   Dwarfs, and Neutron Stars: The Physics of Compact Objects (John Wiley \&
   Sons: New York), 278.
\bibitem{}
   Stoneham, R. J., 1979, J. Phys. A 12, 2187.
\bibitem{}
   Sturner, S. J. 1995, \apj, 446, 292.
\bibitem{}
   Sturner, S.~J. \& Dermer, C.~D., 1994, \apjl, 420, L79.
\bibitem{}
   Sturrock, P. A. 1971, \apj, 164, 529
\bibitem{}
   Thompson. D. J. et al. 1992, \nat, 359, 615.
\bibitem{}
   Thompson, D. J. 1996, in Proc. IAU Colloquium 160
   (Sydney, Australia), in press.
\bibitem{}
   Toll, J. S. 1952, Ph.D. Thesis, Princeton University.
\bibitem{}
   Tsai, W.-Y. \& Erber, T. 1974, \prd, 10, 492.
\bibitem{}
   Usov, V. V. \& Melrose, D. B. 1995, Aust. J. Phys., 48, 571.
\bibitem{}
   Usov, V.~V., \& Shabad, A.~E. 1983, Sov. Astron. Lett., 9, 212.
\bibitem{} 
   Wasserman, I. \& Shapiro, S.L., 1983, \apj, 265, 1036.
\bibitem{}
   Wilke, C. \& Wunner, G. 1996, \prd, submitted.
\bibitem{}
   Wilson, R.~B. et al., 1993, in {\it Isolated Pulsars}, 
   ed. K. van Riper, R.~Epstein \& C. Ho, Cambridge Univ. Press, p. 257.
\bibitem{}
   Wunner, G., Sang, R., \& Berg, D. 1995, \apjl, 455, L51

\end{thebibliography}
\end{document}